\documentclass{vldb}

\usepackage{graphicx}
\usepackage{balance}  

\usepackage[noend,ruled]{algorithm2e} 
\usepackage{balance} 
\usepackage{booktabs} 
\usepackage{color} 
\usepackage{fancyvrb} 
\usepackage{graphicx} 
\usepackage{makecell} 
\usepackage{multirow} 
\usepackage{subfigure} 
\usepackage{url} 
\usepackage{tabulary}
\usepackage{array}
\newcolumntype{L}[1]{>{\raggedright\let\newline\\\arraybackslash\hspace{0pt}}m{#1}}
\newcolumntype{C}[1]{>{\centering\let\newline\\\arraybackslash\hspace{0pt}}m{#1}}
\newcolumntype{R}[1]{>{\raggedleft\let\newline\\\arraybackslash\hspace{0pt}}m{#1}}
\usepackage{xspace} 

\newcommand{\hlight}[1]{#1}
\SetKwRepeat{Do}{do}{while}

\newcommand{\mystore}{\textit{ForkBase}\xspace}
\newcommand{\myvalue}{\texttt{FObject}\xspace}
\newcommand{\myvid}{\texttt{uid}\xspace}
\newcommand{\myid}{\texttt{cid}\xspace}
\newcommand{\mytree}{\texttt{POS-Tree}\xspace}

\begin{document}

\pagestyle{empty}


\title{ForkBase: An Efficient Storage Engine for Blockchain and Forkable Applications}

\author{
Sheng Wang{\small$~^{\#}$},
Tien Tuan Anh Dinh{\small$~^{\#}$},
Qian Lin{\small$~^{\#}$},
Zhongle Xie{\small$~^{\#}$},
Meihui Zhang{\small $~^{\dag}$},
\vspace{0.3em}\\
Qingchao Cai{\small$~^{\#}$},
Gang Chen{\small $~^{\S}$},
Wanzeng Fu{\small$~^{\#}$},
Beng Chin Ooi{\small $~^{\#}$},
Pingcheng Ruan{\small$~^{\#}$}
\vspace{0.4em}\\
\fontsize{11}{11}\selectfont\itshape
$~^{\#}$National University of Singapore,
$~^{\dag}$Beijing Institute of Technology,
$~^{\S}$Zhejiang University
\vspace{0.4em}\\
\fontsize{9}{9}\selectfont\ttfamily\upshape
$~^{\#}$\{wangsh,dinhtta,linqian,zhongle,caiqc,fuwanzeng,ooibc,ruanpc\}@comp.nus.edu.sg,
\vspace{0.3em}\\
\fontsize{9}{9}\selectfont\ttfamily\upshape
$~^{\dag}$meihui\_zhang@bit.edu.cn,
$~^{\S}$cg@zju.edu.cn
\\
}

\maketitle

\begin{abstract}
Existing data storage systems offer a wide range of functionalities to accommodate an equally diverse range of
applications.  However, new classes of applications have emerged, e.g., blockchain and collaborative
analytics, featuring data versioning, fork semantics, tamper-evidence or any combination thereof.  They present new
opportunities for storage systems to efficiently support
such applications
by embedding the above requirements into the storage. 

In this paper, we present \mystore, a storage engine specifically designed to provide efficient support for blockchain and forkable applications.  
By integrating the core application properties into the storage, \mystore not only delivers high performance but also reduces development effort. 
Data in \mystore\ is multi-versioned, and each version uniquely identifies the data content and its history. 
Two variants of fork semantics are supported in \mystore to facilitate any collaboration workflows. 
A novel index structure is introduced to efficiently identify and eliminate duplicate content across data objects.
Consequently,
\mystore is not only efficient in performance, but also in space requirement.
We demonstrate the performance of \mystore using three applications: a blockchain platform, a wiki engine and a
collaborative analytics application. 
We conduct extensive experimental evaluation of these applications against respective state-of-the-art system. 
The results show that \mystore achieves superior performance while significantly lowering the development cost. 

\end{abstract}

\section{Introduction}
\label{sec:intro}

Designing a new application today is made easier by the availability of many storage systems that have
different data models and operation semantics. 
At one extreme, key-value stores~\cite{decandia:2007,
lakshman:2010, web:redis,kemper:2011} provide a simple data model and semantics, but they are highly scalable.
At the other extreme, relational databases~\cite{stonebraker:1987} support more complex, relational models
and strong semantics, i.e. ACID, which render them less scalable. 
In between are systems that make other
trade-offs between data model, semantics and performance~\cite{chang:2006, cooper:2008, bronson:2013,
web:mongodb}.  Despite these many choices, we observe that there emerges a gap between modern applications'
requirements and what existing storage systems have to offer. 

Many classes of modern applications demand properties (or features) that are not a natural fit to current storage
systems. First, blockchain systems such as Bitcoin~\cite{nakamoto:2009}, Ethereum~\cite{web:ethereum} and
Hyperledger~\cite{web:hyperledger}, implement a distributed ledger abstraction --- a globally consistent history of
changes made to some global states. Changes are bundled into blocks, each of which represents a new version of the states.
Because blockchain systems operate in an unstrusted environment, they require the ledger to be {\em tamper evident}, i.e. the
states and their histories cannot be changed without being detected.
Second, collaborative applications, ranging from traditional
platforms like Dropbox~\cite{drago:2012}, GoogleDocs~\cite{web:googledocs}, and Github~\cite{web:github} to more recent
and advanced analytics platforms like Datahub~\cite{maddox:2016} allow many users to work together on the same data.
Such applications need explicit {\em data versioning} to track data derivation history, and {\em fork semantics} to
let users work on independent copies of the data.  
Besides, cryptocurrencies, the most popular blockchain
applications, also allow for temporary forks in the chains. 

Without well-designed storage support for data versioning, fork semantics and tamper evidence, the applications have to
either build on top of systems with partial or no support for these properties, or roll out their own implementations
from scratch. 
Both approaches raise development costs and latency, 
and the former may fail to generalize and may introduce unnecessary
performance overhead. 
One example is that current blockchain platforms (e.g., Ethereum and Hyperledger) build their data structures on top of a key-value store (e.g., LevelDB~\cite{web:leveldb} or RocksDB~\cite{web:rocksdb}). 
The implementations provide tamper evidence, but we observe that they do not always scale well. 
More importantly, they are
not suitable for efficient analytical query processing. 
Another example is collaborative applications over large, relational
datasets, which can be implemented over file-based version control systems such as \texttt{git}. 
However, such implementations do not
scale with large datasets, nor do they support rich query processing. 

Clearly, there are benefits in unifying these properties
and pushing them down into the storage layer. 
One direct benefit is that it reduces development efforts for
applications requiring any combination of these features. 
Another benefit is that it helps applications generalize
better by providing additional features,
such as efficient historical queries, at no extra cost. 
Finally, the storage engine can
exploit performance optimization that is hard to achieve at the application layer. 

We propose a novel and efficient storage engine, called \mystore, that is designed to meet the high demand in
modern applications for versioning, forking and tamper evidence\footnote{\hlight{This is the second version of UStore~\cite{dinh:2017}, which has evolved significantly from the initial design and system.}}.
One challenge in realizing this goal is to keep the storage and computation overhead small when maintaining a large number of data versions. 
Another challenge is to provide small, yet flexible and powerful APIs to various applications.  
Our approach follows well-proven database and system design principles,
and adopts novel designs.
First, a version number uniquely identifies the data content and its history, which can be used to quickly retrieve and verify integrity of the data.
Second,
large objects are split into \textit{data chunks} and organized in a novel index structure, called \mytree, that combines the concepts of content-based slicing~\cite{muthitacharoen:2001}, Merkle tree~\cite{merkle:1987} and
B$^+$-tree~\cite{comer:1979}.  
This structure facilitates efficient identification and removal of duplicate chunks across objects, which
drastically reduces storage overhead especially for incremental data.  
Third, general fork semantics is supported,
providing the flexibility to fork data either implicitly or explicitly.  
The \mytree supports copy-on-write during forking to eliminate unnecessary copies.  
Forth, \mystore offers simple APIs, together with many structured data
types,
which help to reduce development effort and induce a large trade-off space between query
performance and storage efficiency.
Finally, \mystore scales well to many nodes because of a two-layer partitioning scheme which distributes data
evenly, even when the workloads are skew.

\mystore shares some similar goals with recent dataset management systems, namely Decibel~\cite{maddox:2016},
DEX~\cite{chavan:2017} and OrpheusDB~\cite{xu:2017}. 
However, there are two fundamental distinctions. 
First, the other works
target relational datasets.
\mystore's data model is less structured and therefore more flexible. 
Second, the other works are designed mainly for collaborative applications, thus they focus on explicit
data versioning and fork semantics. 
\mystore additionally supports tamper evidence and general fork semantics,
making it useful for blockchain and other forkable applications.

To demonstrate the values of our design, we implement three representative applications on top of \mystore, namely a
blockchain platform, a wiki service, and a collaborative analytics application. 
We observe that only hundreds of lines of code are
required to port major components of these applications onto our system.  
The applications benefit much from the features offered by the storage, especially the fork semantics for
collaborations and tamper evidence for
blockchain. Moreover, as richer semantics are captured in the storage layer, it is feasible to provide
efficient query processing.   
In particular, \mystore enables fast
provenance tracking for blockchain without scanning the whole chain, rendering the blockchain analytics-ready.

In summary, we make the following contributions:
\begin{itemize}

\item We identify common properties in modern applications, i.e., versioning, fork semantics and tamper evidence,  that
are not well addressed and supported  by existing storage systems. 
We examine the benefits of a storage that integrates all of these properties.

\item We design and implement \mystore with the above properties.
It supports generic fork semantics and exposes simple yet elegant APIs.
We propose an index structure for managing large objects, called \mytree,
which is tamper-evident and 
reduces storage overhead for multiple versions of an object.

\item We demonstrate the usability and efficiency of \mystore by implementing three representative and complex applications, namely a blockchain
platform, a wiki service and a collaborative analytics application.
We evaluate the performance of \mystore and the
three applications against their respective state-of-the-art implementations. 
We show that \mystore improves these applications in
terms of coding complexity, storage overhead and query efficiency. 
\end{itemize}

In the following, we first motivate the design of \mystore in Section~\ref{sec:feature}.  We then present its
data model and APIs in Section~\ref{sec:api}, and the detailed design and implementation in Section~\ref{sec:design}.
We describe the implementation and evaluation of three applications in Section~\ref{sec:app}
and~\ref{sec:exp} respectively, and conclude in Section~\ref{sec:conclu}.

\section{Background and Motivations}
\label{sec:feature}
In this section, we discuss three properties underpinning many modern applications and related systems that are related to
\mystore.  

\subsection{Data Versioning}
Data versioning is an important concept in applications that keep track of data evolution history, in which any update
made on the data results in a new copy (or version). 
The history can be either linear or non-linear. Systems supporting linear
data versioning include multi-versioned file systems~\cite{santry:1999, strunk:2000, soules:2003} and temporal databases (e.g., relational~\cite{ahn:1986, salzberg:1999, tansel:1993}, graph~\cite{khurana:2013} and array~\cite{seering:2012}). Non-linear data versioning systems can support file-based, unstructured data such as version control systems
(e.g., \texttt{git}, \texttt{svn} and \texttt{mercurial}), or more structured data such as
Decibel~\cite{maddox:2016}, DEX~\cite{chavan:2017} and OrpheusDB~\cite{xu:2017}. Blockchain is another example of
data versioning systems, in which each block represents a version of the global states. 

One challenge in the support of data versioning is to reduce storage consumption.  The most common approach, called {\em
delta-based deduplication},  is to store only the differences (or deltas) across data versions, and reconstruct the content
from a chain of deltas.  Decibel proposes several physical data layouts for storing deltas, while
OrpheusDB bolts on a relational database in order to take advantage of the latter's
query functionalities.  The trade-off between storage and reconstruction cost has been studied
in~\cite{bhattacherjee:2015}.  Another approach to storage reduction is {\em content-based deduplication}. File
systems~\cite{paulo:2014} and \texttt{git}, for examples, adopt this approach to eliminate coarse-grained duplicates. In
particular, data is split into files or chunks, each of which is uniquely identified by its content. The systems then
detect and eliminate all duplicates. We note that both deduplciation techniques can be combined. For example,
\texttt{git} employs content-based technique at the file level, and delta-based technique for linked versions during the
\texttt{repack} process.

\mystore applies content-based deduplication at the chunk level.
Compared to similar deduplication technique
in file systems which uses large chunk sizes and treats the file content as unstructured data, \mystore uses smaller
chunks and splits the data based on its structure. For instance, a list object containing multiple elements is only
split at element boundaries, thus avoiding the need to reconstruct an element from multiple chunks. Taking the structure of
data object into account makes updates and dedpulciations more efficient.
\hlight{
Noms~\cite{web:noms} applies chunk-level deduplication similar to \mystore.
However, it targets single storage instance with fast synchronization,
whereas \mystore applies deduplication over multiple storage instances to
optimize for large-volume data accesses and modifications.
}
Compared to the delta-based technique used in
Decibel to remove duplicates within a dataset, \mystore achieves better storage reduction because it can also eliminate
cross-dataset duplicates generated by uncoordinated teams\footnote{Like any other content-based techniques, the
deduplication is less effective than delta-based techniques when the deltas are much smaller than the chunks.}. 
\hlight{
Furthermore,
\mystore offers richer branch management (discussed below) to support more diverse collaborative workflows.  
}

\subsection{Fork Semantics}
\label{subsec:fork_app} 

Fork semantics elegantly captures non-linearity of the data evolution history.  It consists of two core operations:
fork and conflict resolution.  A fork operation creates a new data \emph{branch}, which evolves independently and its
local modifications are isolated from other branches'.  Data forks isolate conflicted updates which can then be merged
via the conflict resolution operation.  Applications exploiting this semantics can be divided into two categories: one
that invokes \emph{on-demand} (or explicit) forks and the other that relies on \emph{on-conflict} (or implicit) forks.

On-demand forks are found in applications that have explicit demand for isolated (or private) branches. Source code
version control systems like \texttt{git} allow users to fork a new branch for their own development and only merge
changes to the main codebase after they are well tested. Similarly, collaborative analytics applications such as  
Datahub~\cite{bhardwaj:2015} allow branching off from an original dataset before applying transformation to the data,
e.g., data cleansing, correction and integration.
On-conflict forks are used in applications that implicitly fork a state upon
concurrent modifications of the same data.  
Transactional systems with weak consistency, e.g.
TARDiS~\cite{crooks:2016}, fork the database state during the concurrent execution of conflicting transactions, and delay
(user-defined) conflict resolution.
In blockchain applications, for instance
Bitcoin~\cite{nakamoto:2009} and Ethereum~\cite{web:ethereum},
forks arise when multiple blocks are appended at the same time to an old block.
They are resolved by taking the longest chain or by more complex mechanisms like GHOST~\cite{ghost}. 

\mystore is thus motivated to be the first system to natively support both implicit and explicit forks.
To facilitate application development, it also provides a number of built-in conflict resolution operations.

\subsection{Tamper Evidence}

Security conscious applications demand protection against malicious modifications, not only from
external attackers but also from malicious insiders.  One example is outsourced services like
storage~\cite{kallahalla:2003} or file system~\cite{li:2004}, which provide mechanisms to detect tampering (forking) of
the update logs.  Another example is blockchain platforms~\cite{nakamoto:2009, kemper:2011, web:ethereum}, which require
tamper evidence for the ledger. The blockchain combines the tamper-evident ledger with a distributed consensus protocol
to ensure that the global states are immutable across multiple nodes. We note that there is an increasing demand for
performing analytics on blockchain data~\cite{morgan16,chainalysis,blocksci}, which existing blockchain storage engines were not
designed for.  
Specifically, current systems implement the ledger on top of a key-value storage.
Further, they focus on tamper evidence and do not efficiently support querying the states' history. 

\mystore provides tamper evidence against malicious storage providers. Given a version number, the application can fetch
the corresponding data from the storage provider and verify whether the content and its history have been changed.
All data objects in \mystore are tamper-evident, and hence can be leveraged to build better data models for blockchain.
In particular, the blockchain's key data structures implemented on top of \mystore are now easy to maintain without
incurring any performance overhead.  
Furthermore, the richer structured information captured in \mystore makes the
blockchain analytics-ready.

\subsection{Design Overview}

\begin{figure}
  \centering
  \includegraphics[width=0.48\textwidth]{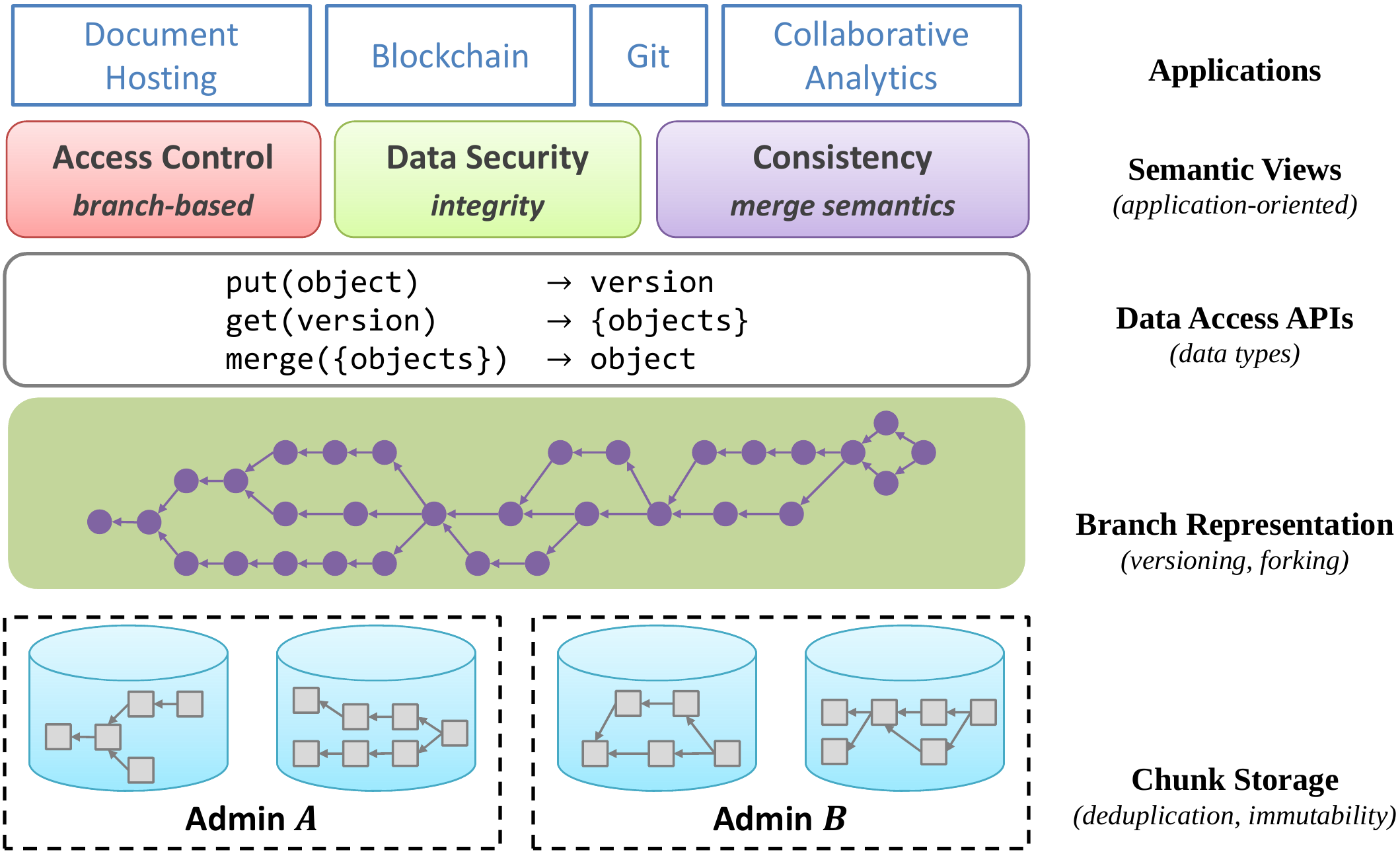}
  \caption{The \mystore stack offers advanced features to various classes of modern applications.}
  \label{fig:stack}
\vspace{-1ex}
\end{figure}

Figure~\ref{fig:stack} shows the \mystore's stack, 
illustrating how the storage unifies the common properties and adds
values to modern applications.  
At the bottom layer, data is chunked and deduplicated.
At the representation
layer, versions and branches are organized in such a way that enables tamper evidence and efficient tracking of the version
history. 
The next layer exposes APIs that combine general fork semantics and structured data types. Other features such
as access control and customized merge functions, can be added to the view layer to further enrich the top-layer
applications.

\section{Data Model and APIs}
\label{sec:api}
In this section, we describe the data model and basic operations, and show an example on how to leverage core features from provided APIs.

\subsection{FObject}

\mystore extends the basic key-value data model: 
each data object in \mystore is identified by
a {\em key}, and contains a {\em value} of a specific {\em type}. For each key, it is possible to retrieve not
only its latest value, but also its evolution history.  
Similar to other data versioning systems, \mystore
organizes versions in a directed acyclic graph (DAG) called \emph{object derivation graph}.
Each node in the graph is a structure called \myvalue, and it is associated with a unique identifier \myvid.
Links between {\tt FObjects} represent their derivation relationships.

\begin{figure}
\centering
\begin{verbatim}
struct FObject {
  enum type;  // object type
  byte[] key;  // object key
  byte[] data;  // object value
  int depth;  // distance to the first version
  vector<uid> bases;  // versions it derives from
  byte[] context;  // reserved for application
}
\end{verbatim}
\vspace{-2em}
\caption{The \myvalue structure.}
\label{fig:object}
\end{figure}

The structure of a \myvalue is shown in Figure~\ref{fig:object}.  
The {\tt context} field is reserved for
application metadata, for examples commit message in git, or nonces value for blockchain
proof-of-work~\cite{dwork:1992}. 
Access to a \myvalue is 
via the {\tt Put} and {\tt Get} APIs listed
in Table~\ref{tab:api}. In particular: 
\begin{itemize}
  \item {\tt Put(key, <branch>, value)} - write a new value to the specified branch. When {\tt branch} is
  absent, write to the {\em default branch}.
  \item \texttt{Get(key, <branch>)} - read the latest value from the specified branch. When {\tt branch} is
  absent, read from the {\em default branch}. 
\end{itemize}
It can be seen that the \mystore 's data model is compliant to the basic key-value model when only the default branch is used. 

\subsection{Tamper-Evident Version}
\label{subsec:version}
Each \myvalue is associated with a \myvid representing the data version. An important property of the \myvid is that
it is tamper-evident. The \myvid uniquely identifies both the object value and its derivation history. Two {\tt
FObjects} are considered logically equivalent, i.e. having the same \myvid, only when they have the same value
and derivation history.  
Suppose the application is given $v_l$ as the latest version of an object, let $V =
\langle v_1, v_2, .., v_l\rangle$ be the derivation history. The storage cannot prove to the application
that a version $v' \notin V$ is part of the object history. In other words, the storage cannot tamper with the
object value and its history. 

\mystore realizes this property by linking versions via a cryptographic hash chain. In particular, each
\myvalue stores the hashes of the previous versions it derives from in the {\tt bases} field.  
Two important operations on versions are supported, namely {\tt Diff} and {\tt LCA}. 
The former returns the
differences between two {\myvalue}s of the same types (they could be of different keys). 
The latter
returns the least common ancestor of two {\myvalue}s with the same key.

\subsection{Fork and Merge}
\label{subsec:fork}

\begin{table}
  \centering  
  \caption{\mystore APIs.}
  \label{tab:api}
  \vspace{1ex}  
  \begin{tabular}{c|c|cc|c}
    \hline
     & Method & FoD & FoC & Ref \\
    \hline
    \multirow{2}{*}{\texttt{Get}}
      & \texttt{Get(key,branch)} & $\checkmark$ & & M1 \\
      & \texttt{Get(key,uid)} & $\checkmark$ & $\checkmark$ & M2 \\
    \hline
    \multirow{2}{*}{\texttt{Put}}
      & \texttt{Put(key,branch,value)} & $\checkmark$ & & M3 \\
      & \texttt{Put(key,base\_uid,value)} & & $\checkmark$ & M4 \\
    \hline
    \multirow{3}{*}{\texttt{Merge}}
      & \texttt{Merge(key,tgt\_brh,ref\_brh)} & $\checkmark$ & & M5 \\
      & \texttt{Merge(key,tgt\_brh,ref\_uid)} & $\checkmark$ & & M6 \\
      & \texttt{Merge(key,ref\_uid1,...)} & & $\checkmark$ & M7 \\
    \hline    
    \multirow{3}{*}{\texttt{View}}
      & \texttt{ListKeys()} & $\checkmark$ & $\checkmark$ & M8 \\
      & \texttt{ListTaggedBranches(key)} & $\checkmark$ & & M9 \\
      & \texttt{ListUntaggedBranches(key)} & & $\checkmark$ & M10 \\
    \hline    
    \multirow{4}{*}{\texttt{Fork}}
      & \texttt{Fork(key,ref\_brh,new\_brh)} & $\checkmark$ & & M11 \\
      & \texttt{Fork(key,ref\_uid,new\_brh)} & $\checkmark$ & & M12 \\
      & \texttt{Rename(key,tgt\_brh,new\_brh)} & $\checkmark$ & & M13 \\
      & \texttt{Remove(key,tgt\_brh)} & $\checkmark$ & & M14 \\
	\hline   
	\multirow{3}{*}{\texttt{Track}}
	  & \texttt{Track(key,branch,dist\_rng)} & $\checkmark$ & & M15 \\
      & \texttt{Track(key,uid,dist\_rng)} & & $\checkmark$ & M16 \\
      & \texttt{LCA(key,uid1,uid2)} & $\checkmark$ & $\checkmark$ & M17 \\
    \hline
\end{tabular}
\end{table}

The latest version of a branch is called the branch {\em head}. A branch is only modifiable
at the head. However, to change a historical version, a new
branch can be created (forked out) at that version to make it modifiable. There are no restrictions on the
number of branches per key. \mystore generalizes fork operations by providing two fork semantics: \emph{fork
on demand} (FoD) and \emph{fork on conflict} (FoC). Table~\ref{tab:api} details the semantics for the basic
operations supported in \mystore. 

\subsubsection{Fork on Demand}

\begin{figure}
\centering
\begin{minipage}{.45\textwidth}
\centering
\subfigure[]{
  \includegraphics[scale=0.6]{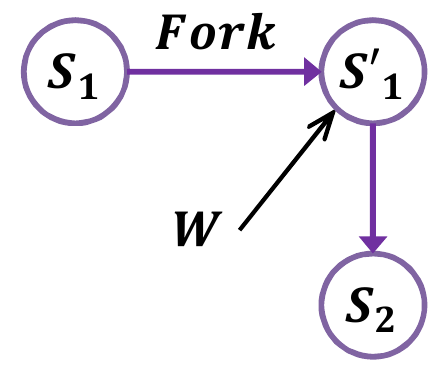}}
\hspace{2em}
\subfigure[]{
  \includegraphics[scale=0.6]{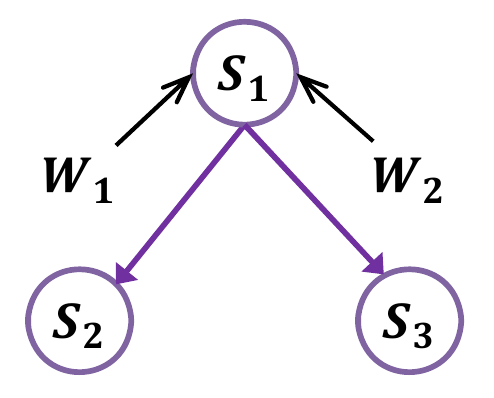}}
\caption{Generic fork semantics supported for both (a) fork on demand and (b) fork on conflict.}
\label{fig:fork}
\end{minipage}
\vspace{-1ex}
\end{figure}

A branch is created explicitly before any modifications.  For example, in
Figure~\ref{fig:fork}(a) a branch with head version $S_1$ is forked to create a new branch.  Then a new
update $W$ is applied to the new branch creating a new version $S_2$. $S_2$ is now the head of an independent
branch.  Branches generated in this way require user-defined names, and thus we refer to them as
\emph{tagged} branches. The \texttt{Fork} operation creates a new tagged branch by taking as input an existing
branch (M11 in Table~\ref{tab:api}), or a non-head \myvalue of the existing branch (M12). (M9) lists all
branch names and their head {\myvid}s. (M1) and (M3) allow for reading and modifying the head version of a
given branch. Non-head versions can be read using (M15).

\subsubsection{Fork on Conflict}

Branches are implicitly created from concurrent, conflicting \texttt{Put} (M4)
operations in which different changes are made to the same version. For example, in Figure~\ref{fig:fork}(b)
two updates $W_1$ and $W_2$ are applied to the head version $S_1$ concurrently.  
This is common in
decentralized environments where concurrent updates from remote users may not be immediately visible. 
The
result is that two branches with different head versions $S_2$ and $S_3$ are created. Branches generated in
this way can only be identified by their {\myvid}s, and thus we refer to them as \emph{untagged} branches.
Conflicting branches can be checked using (M10) which  returns a single head version if no conflict is found.
Otherwise, all conflicting head versions are returned, with which the application can decide when and how
the conflicts should be resolved.

\subsubsection{Merge}

A tagged branch can be merged with another tagged branch (M5) or with a specific version (M6).  In both cases,
only the first branch's head is updated such that the new head contains data from both branches. A collection
of untagged branches can be merged using (M7), after which the input branches are logically
replaced  with a new branch.  When conflicts are detected during a merge, the application can resolve
them in many ways (Section~\ref{subsec:branch}). To simplify the merge process, \mystore provides
type-specific merge functions for the built-in data types.

\begin{figure}
\centering
\scriptsize
\begin{Verbatim}[frame=single]
ForkBaseConnector db;
// Put a blob to the default master branch
Blob blob {"my value"};
db.Put("my key", blob);
// Fork to a new branch
db.Fork("my key", "master", "new branch");

// Get the blob
FObject value = db.Get("my key", "new branch");
if (value.type() != Blob)
  throw TypeNotMatchError;
blob = value.Blob();

// Remove 10 bytes from beginning and append new
// Changes are buffered in client
blob.Remove(0, 10);
blob.Append("some more");
// Commit changes to that branch
db.Put("my key", "new branch", blob);
\end{Verbatim}
\caption{Fork and modify a \texttt{Blob} object.}
\label{fig:blob}
\vspace{-1ex}
\end{figure}

\subsection{Data Type}
\label{subsec:types}
\mystore provides many built-in, structured data types. They can be categorized into two classes:
\emph{primitive} types and \emph{chunkable} types.

\textbf{Primitive} types include simple types such as
\texttt{String}, \texttt{Tuple} and \texttt{Integer}. They are small-size objects that are optimized for fast
access. These objects are not deduplicated, since the benefits of sharing small data does not offset the extra
overheads of deduplication. Apart from the basic \texttt{Get} and \texttt{Set} operations, type-specific
operations are provided to modify primitive objects. Examples include \texttt{Append} and \texttt{Insert} for
\texttt{String} and \texttt{Tuple} types, and \texttt{Add} and \texttt{Multiply} for numerical types.

\textbf{Chunkable} types are complex data structures, for examples \texttt{Blob}, \texttt{List}, \texttt{Map} and
\texttt{Set}.  A chunkable object is stored as a {\mytree} and deduplicated (Section~\ref{subsec:tree}). 
The chunkable object is most suitable to represent data that grows fairly large due to many updates, but each update
touches only a small portion of the data. In other words, a new version has significant overlap with the previous
version. The read operation returns only a handler, while the actual data is fetched gradually on demand.
Iterator interfaces are provided to efficiently traverse large objects.

The rich collection of built-in types makes it easy to build high level structures, such as relational
tables (Section~\ref{sec:app}).
Note that different data types may have similar logical representation but
different performance, for example \texttt{String} and \texttt{Blob}, or \texttt{Tuple} and \texttt{List}.
The application 
is able to choose those types are more suitable based on their expected workloads. 

\subsection{Example}

In \mystore, data can be manipulated at two granularities,
i.e., at an individual object, and at a branch of objects.
\mystore exposes easy-to-use interfaces that
combine both object manipulation and branch management.
Figure~\ref{fig:blob} shows an example of forking and editing a {\tt Blob} object.
Since \texttt{Put} serves for both insertion and update,
the $value$ input to
the \texttt{Put} operation could be either a whole new object or the base object that has undergone a
sequence of updates. When multiple updates of the same object are batched, \mystore only retains the final version.

\section{Design and Implementation}
\label{sec:design}

\label{sec:arch}

\begin{figure}
  \centering
  \includegraphics[width=0.48\textwidth]{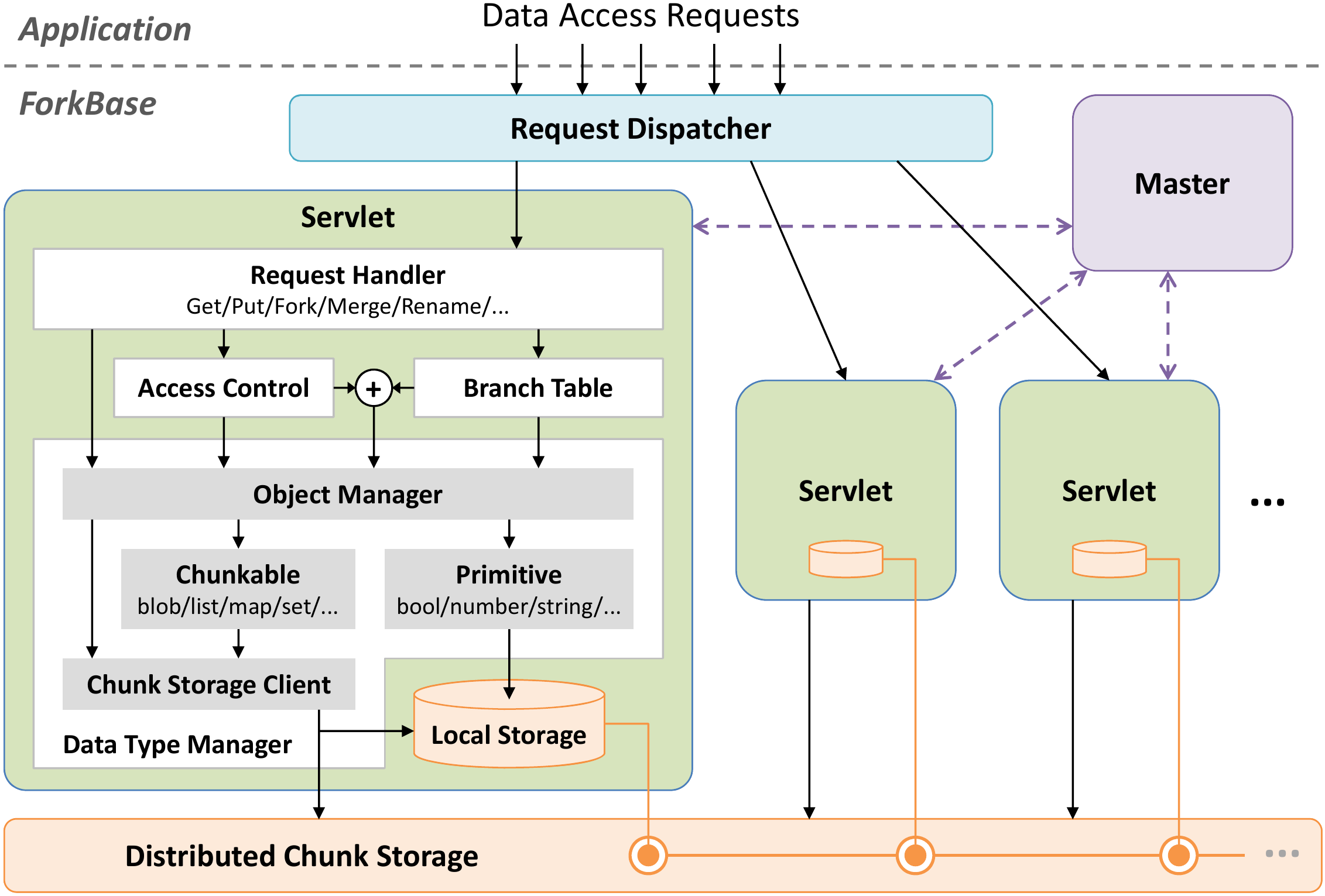}
  \caption{Architecture of a \mystore cluster.}
  \label{fig:arch}
\vspace{-1ex}
\end{figure}

In this section, we present the detailed design and implementation of \mystore.
The system can used as an embedded storage or run as a distributed service. 

\subsection{Architecture}
Figure~\ref{fig:arch} shows the stack of a \mystore cluster consisting of four main components: a
\emph{master}, a \emph{dispatcher}, a \emph{servlet} and a \emph{chunk storage}.  When used as an embedded
storage, only one servlet and one chunk storage are instantiated.  The \emph{servlet} executes requests using
three sub-modules. First, the \emph{access controller} verifies
request permission before execution. Second, the \emph{branch table} maintains branch heads for both tagged
and untagged branches. Third, the \emph{object manager} handles object manipulations, hiding the physical
data representation from the main execution logic. The \emph{chunk storage} persists and provides access to
data chunks.  When deployed as a distributed service, the \emph{master} maintains the cluster runtime
information,
while the \emph{request dispatcher} receives and forwards requests to the corresponding servlets.  Each
servlet manages a disjoint subset of the key space, as determined by a routing policy.  All chunk storage
instances form a large pool of storage, which is accessible by any remote servlets. In fact, each servlet is
co-located with a local chunk storage which enables fast data access and persistence. 

\subsection{Physical Data Representation}
\label{subsec:represent}
\mystore objects are stored in the form of data chunks of various lengths.  A small and simple object, i.e. of
primitive types, contains a single chunk. A large and complex object, i.e. of chunkable types, comprises multiple
chunks organized as a \mytree.

\begin{table}
  \centering
  \caption{Chunk Content.}
  \label{tab:chunk}
  \vspace{1ex} 
  \begin{tabular}{c|l}
    \hline
    Type & Content \\
    \hline
    \texttt{Meta} & metadata for an \myvalue \\
	\texttt{UIndex} & index entries for unsorted types (\texttt{Blob}, \texttt{List})\\    
    \texttt{SIndex} & index entries for sorted types (\texttt{Set}, \texttt{Map})\\
    \texttt{Blob} & a sequence of raw bytes \\
    \texttt{List} & a sequence of elements \\
    \texttt{Set} & a sequence of sorted elements \\
    \texttt{Map} & a sequence of sorted key-value pairs \\
    \hline
\end{tabular}
\end{table}

\subsubsection{Chunk and cid}

A \emph{chunk} is the basic unit of storage in \mystore. There are multiple chunk types
(Table~\ref{tab:chunk}), each corresponding to a chunkable data type.
A chunk is uniquely identified by its
\myid which is computed from the content contained in the chunk:
$$ chunk.\texttt{cid} = H(chunk.bytes)$$
where $H$ is a cryptographic hash function (e.g., SHA-256, MD5)
taking raw bytes of a chunk as input.
Due to the property of cryptographic hashes, each chunk will have a unique \myid,
i.e., chunks with the same \myid should contain identical content.
\mystore uses SHA-256 as the default hash
function,
but faster alternatives, e.g., BLAKE2,
can also be used to reduce computational overhead.
The chunks are stored in chunk storage and can be accessed via {\myid}s.

\subsubsection{FObject and Data Types}
Recall that a \texttt{Get} request returns a \myvalue, whose layout is 
shown in Figure~\ref{fig:object}. A serialized chunk of a \myvalue is called a
{\em meta} chunk.  The \myvalue's \myvid is in fact an alias for the meta chunk's \myid. For a \myvalue of
primitive type, the chunk content is embedded in the meta chunk's \texttt{data} field to facilitate fast access.
For chunkable type object, its meta chunk only contains a \myid in the \texttt{data} field, which points to
an external data structure, i.e. the \mytree.  As a result, updates to a chunkable object only modify the
\myid value in the \myvalue structure.

\subsection{Pattern-Oriented-Split Tree}
\label{subsec:tree}
Large structured objects are not usually accessed in their entirety. Instead, they require fine-grained
access, such as element look-up, range query and update. These access patterns require index structures,
e.g., $B^+$-tree, to be efficient. However, existing index structures are not suitable in our context that has
many versions and the versions can be merged. For
example, $B^+$-trees and variants that support branches~\cite{jiang:2000}, use capacity-based splitting
strategies, and their structures are determined by the values being indexed and by the insertion order. For
example, inserting value $A$ followed by $B$ may result in a different B$^+$-tree to inserting $B$
followed by $A$. There are two consequences when maintaining many versions. First, it is difficult to share (i.e., deduplicate) both index and leaf nodes
even when two trees contain the same elements. Second, it is costly to find the differences between two versions and
merge them, because of the structural differences. One simple solution is to have fixed-size nodes, which
eliminates the effect from insertion order.  However, such an approach introduces another issue, called
boundary-shifting problem~\cite{eshghi:2005}, when an insertion occurs in the middle of the structure.

\begin{figure}
  \centering
  \includegraphics[width=0.45\textwidth]{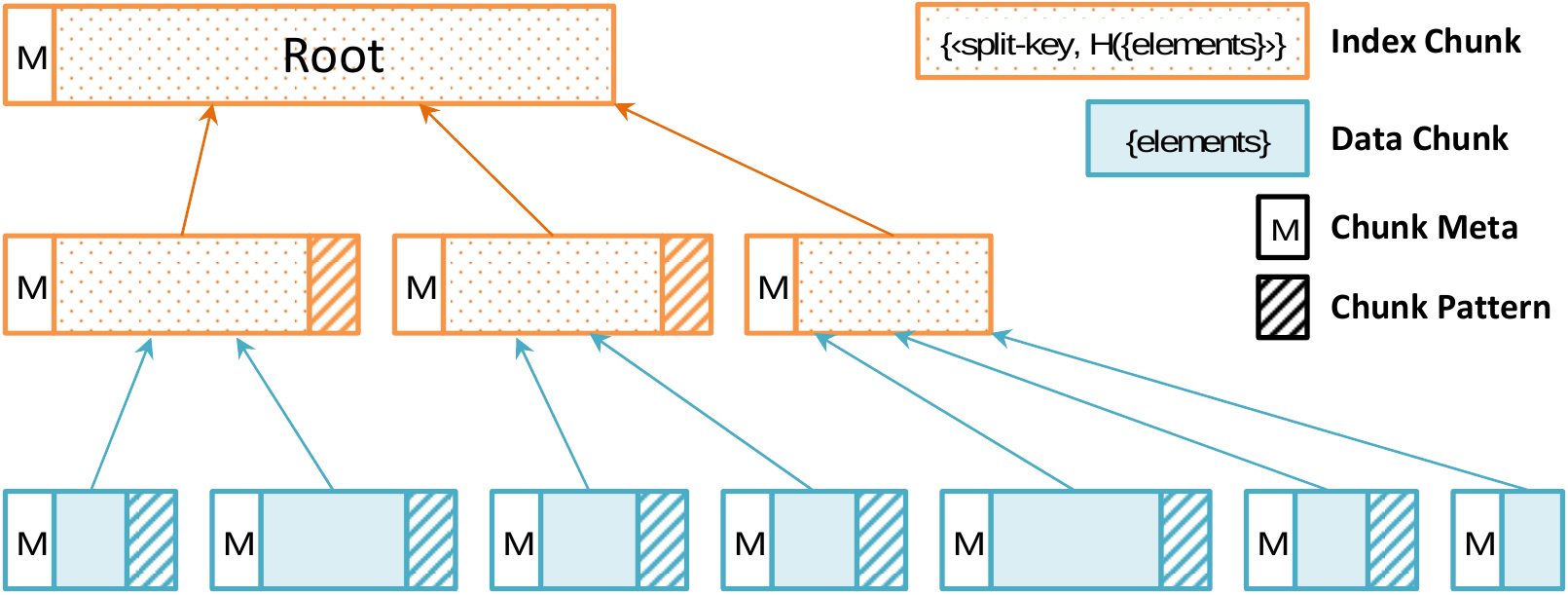}
  \caption{Pattern-Oriented-Splitting Tree (POS-tree) resembling a B$^+$-tree and Merkle tree.}
  \label{fig:pos-tree}
\end{figure}

To solve above issues, we propose a novel index structure, called \textit{Pattern-Oriented-Split Tree} (\mytree), which has the
following properties: 
\begin{itemize}
\item It is fast to look up and update elements;
\item It is fast to find differences and merge two trees;
\item It is effective in deduplication;
\item It provides tamper evidence.
\end{itemize}
This structure is inspired by content-based slicing~\cite{muthitacharoen:2001}, and resembles a combination of
a B$^+$-tree and a Merkle tree~\cite{merkle:1987}. In \mytree, the node (i.e., chunk) boundary is
defined as patterns detected from the object content. Specifically, to construct a node, we scan from
the beginning until a pre-defined pattern occurs, and then create a new node to hold the scanned content.
Because the leaf nodes and internal nodes have different degrees of randomness, we define different patterns
for them. In following, we first describe the basic tree structure, and then discuss how it is constructed.

\subsubsection{Tree Structure}
Figure~\ref{fig:pos-tree} illustrates the structure of a \mytree.  Each node is stored as a chunk. Index nodes
are stored as \texttt{UIndex} or \texttt{SIndex} chunks, whereas leaf node chunks are of the object types,
such as \texttt{Blob}, \texttt{List} or \texttt{Map} chunks,
as listed in Table~\ref{tab:chunk}.  Similar to B$^+$-trees, an index
node contains a number of entries for its child nodes.  Each entry consists of a child node's \myid and the
corresponding split key (for \texttt{SIndex} or element count for \texttt{UIndex}).  To look up a specific
key (or a position for \texttt{UIndex}), we adopt the same strategy as in B$^+$-trees, i.e. follow the path
guided by the split keys. Therefore, accessing a chunkable object is efficient because only the relevant nodes
are fetched instead of the entire tree. \mytree is a Merkle tree in the sense that the child nodes' {\myid}s
are cryptographic hashes of their content.
Hence, two objects with identical data will have the same {\mytree}.
In addition, comparing two trees can be done efficiently by recursively comparing the \myid{s}. 
 
\subsubsection{Splitting a Leaf Node}
\label{subsubsec:chunking}

Here we describe the pattern used to split a leaf node. Given a $k$-byte sequence $(b_1,..,b_k)$, let $P$ be a function
taking $k$ bytes as input and returning a pseudo-random integer of at least $q$ bits. The pattern is said to occur if
and only if:
$$ P(b_1 ... b_k) \ \text{BITWISE\_AND} \ (2^q - 1) = 0 $$
In other words, the pattern occurs when the function $P$ returns $0$ for the $q$ least significant bits. The
complexity of $P$ is at least $O(k)$.  We use a special class of hash function, called rolling hash, that
supports fast computation over sequence windows (e.g., Rabin-Karp, cyclic polynomial, and moving sum).  In
particular, we implement $P$ as
the \emph{cyclic polynomial}~\cite{cohen:1997} rolling hash function, which is of the form:
$$P(b_1 ... b_k) = s^{k-1}(h(b_1)) \oplus s^{k-2}(h(b_2)) \oplus ... \oplus s^{0}(h(b_k))$$ 
where $\oplus$ is exclusive-or operator, and $h$ maps a byte to an integer in $[0, 2^q)$.  $s$ is a
function that shifts its input by 1 bit to the left, and then pushes the $q$-th bit back to the lowest position. 
This function can be computed recursively as follows: 
$$P(b_1 ... b_k) = s(P(b_0 ... b_{k-1})) \oplus s^k(h(b_0)) \oplus s^0(h(b_k))$$
Each time, it removes the oldest byte from the active set and adds the new one.
As a result, the computation cost amortizes with many sequence windows of $k$ bytes. 

Initially, the entire object content is treated as one byte sequence. The pattern detection process scans this
sequence from the beginning. When a pattern occurs, a leaf node is created with the recently scanned data.  If
a pattern occurs in the middle of an element (e.g., a key-value pair in \texttt{Map}), the chunk boundary is
extended to cover the whole element, so that no elements are stored in more than one chunks.
In this way, each
leaf node (except for the last node) ends with a pattern, as shown in Figure~\ref{fig:pos-tree}.

\begin{algorithm}[t]
\caption{\mytree Construction}
\label{alg:construct}
\KwIn{a list of data elements $data$}
\KwOut{\myid of constructed \mytree 's root}
PatternDetector $hash$\;
List$<$Element$>$ $elements$, $new\_entries$\;
Chunk $chunk$\;
cid $last\_commit$\;
$new\_entries$ = $data$\;
\tcc{use pattern function P for leaf nodes}
$hash$ = new P()\;
\Do{$new\_entries$.size() $>$ 1}{
	move all elements in $new\_entries$ to $elements$\;
	\For{each $e$ in $elements$}{
		$chunk$.append($e$)\;
		feed $e$ into $hash$ to detect pattern\;
		\If{detected pattern or exceeded max length}{
			$last\_commit$ = $chunk$.commit()\;
			add index entry of $chunk$ into $new\_entries$\;
		}
	}
	\tcc{last chunk may not have pattern}	
	\If{$chunk$ is not empty}{
		$last\_commit$ = $chunk$.commit()\;
		add index entry of $chunk$ into $new\_entries$\;
	}
	\tcc{use pattern function P' for index nodes}
	$hash$ = new P'()\;
\tcc{loop until root is found}
}
\Return{$last\_commit$;}
\vspace{-1ex}
\end{algorithm}

\subsubsection{Splitting an Index Node}
The rolling hash function used for splitting leaf nodes has good randomness which keeps the structure
balanced against skewed application data. 
However, we observe that it is costly: it accounts for $20\%$ of the cost for building
\mytree{s}. 
Thus, for index nodes, we use a more efficient function $P'$ that exploits the randomness
of the leaf nodes' {\myid}s. 
In particular, $P'$ directly takes the \myid and determines that a pattern
occurs when: 
$$\texttt{cid} \ \text{BITWISE\_AND} \ (2^r - 1) = 0$$ 
When a pattern is detected, all scanned index entries are stored in a new index node. 
This process is repeated
for upper layers until reaching the root node. Algorithm~\ref{alg:construct} demonstrates the bottom-up construction of a new \mytree. 

When updating an existing \mytree,
only affected nodes are reconstructed,
which results in a copy-on-write strategy.
A node splits when a new pattern is found in between.
When an existing boundary pattern changes, the next node needs to be merged. However, no subsequent chunks are
involved during the reconstruction, because the boundary pattern of the last merged chunk is preserved.
Since $P'$ limits the pattern inside a single index entry, it reduces the chance that existing patterns are
changed (compared to using $P$ where $k$ is larger than the \myid length). 

The expected node size (i.e., chunk size) can be configured by setting the values of $q$ and $r$.
By default, \mystore applies a pre-defined chunk size (e.g., 4 KB) for all nodes,
but it is beneficial to configure type-specific chunk sizes.  For example, \texttt{Blob}
chunks storing multimedia data can have large sizes, whereas \texttt{Index} chunks may need smaller sizes
since they only contain tiny index entries.  To ensure that a node will not grow infinitely large, an
additional constraint is enforced: the chunk size cannot be $\alpha$ times bigger than the average size;
otherwise it is forcefully chunked.  Therefore, the probability of forced chunking is equal to
$(1/e)^{\alpha}$, which can be set very low (e.g $0.0335\%$ when $\alpha = 8$).

\hlight{
We note that \mytree is not designed for cases in which the object content is a sequence of repeated items (or
bytes). Since there is no pattern in the leaf nodes, all nodes have the maximum chunk size.  Consequently, an
insertion in the middle leads to boundary shift, thus incurring overhead for re-splitting the node.
Nevertheless, \mytree remains deduplicatable, because the leaf nodes are identical and can be deduplicated.
}

\hlight{
\subsection{Chunk Storage}
\label{subsec:storage}
The \emph{Chunk storage} manages data chunks and exposes a key-value interface. The key is a \myid,
while the value is that chunk of raw bytes. With tamper evidence at the chunk level, the chunk storage can verify
  \texttt{Get-Chunk} and \texttt{Put-Chunk} requests on demand.
When \texttt{Put-Chunk} request contains an existing \myid, the storage can
respond immediately. This is thanks
to the deduplication mechanism such that the same chunk from previous request can be reused. Since chunks are immutable, a
log-structured layout is used for persistence, which also facilitates fast retrieval of consecutively
generated chunks in a \mytree. To improve data durability and fault tolerance, chunks
can be replicated over multiple nodes (or chunk storage instances). Such replication does not significantly
affect the deduplication; there are only $k$ copies of any chunk in the storage. Furthermore, replicas help
reduce the latency of data access,
e.g., by placing a replica on the servlet that frequently accesses its data.  
%
%
}


\subsection{Branch Management}
\label{subsec:branch}
For each data key there is a \emph{branch table} that
holds all its branches' heads, i.e., the latest \myid{s} of the branches.  The branch table comprises two data
structures for tagged and untagged branches respectively.  Tagged branches are maintained in a map
structure called \emph{TB-table}, in which each entry consists of a tag (i.e., branch name) and a head \myid.
Untagged branches are maintained in a set structure called \emph{UB-table}, in which each entry is simply a
head \myid. UB-table essentially maintains all the leaf nodes in the object derivation graph.

\subsubsection{Branch Update}

The TB-table is updated during the \texttt{Put-Branch} operation (M3).  Once the new \myvalue is constructed,
its \myid replaces the old branch head in the table.  The \texttt{Fork-Branch} operation (M11) simply creates a
new entry pointing to the referenced branch head.  Similarly, operations (M12-M14) only modify entries in the
TB-table, without creating new objects.
Concurrent updates on a tagged branch
are serialized by the servlet.
Moreover,
to further protect from overwriting others' changes by accident,
additional \emph{guarded} APIs are provided,
such as:
$$\texttt{PUT(key, branch, value, guard\_uid)}$$
which ensures that the {\tt Put} is successful only if the current branch head is equal to {\tt guard\_vid}.

The UB-table is updated whenever a new \myvalue is created.  Once the \myvalue is constructed, its \myid is
added to the UB-table, and its base \myid is removed from the table.  When the base \myid is not found
in the table, it means the base version has already been derived by others.
If the new \myvalue already exists in the storage (i.e., from previous equivalent operations), the UB-table
simply ignores it.

\subsubsection{Conflict Resolution}
\label{subsubsec:merge}

A three-way merge algorithm is used for \texttt{Merge} (M5-M7) operations.  To merge two branch heads $v_1$ and
$v_2$, three versions ($v_1$, $v_2$ and \texttt{LCA(}$v1,v2$\texttt{)}, i.e., the most recent version where
they start to fork) are fed into the merge function.  If the merge fails, it returns a conflict list,
calling for conflict resolution. This can be done at the application layer and the merged result sent back to the
storage. Simple conflicts can be resolved using built-in resolution functions (such as \texttt{append}, \texttt{aggregate} and
\texttt{choose-one}). \mystore allows users to hook customized resolution strategies which are executed upon
conflicts.

\subsection{Cluster Management}

When deployed as a distributed service, \mystore uses a \emph{hash-based two layer partitioning} that
distributes workloads evenly among nodes in the cluster:
\begin{itemize}
\item \textbf{Request dispatcher to servlet:} requests received by a dispatcher are partitioned and sent to
the corresponding servlet based on the request keys' hash.
\item \textbf{Servlet to chunk storage:} chunks created in a servlet are partitioned
based on {\myid}s, and then forwarded to the corresponding chunk storage. Meta chunks are always
stored locally.
\end{itemize}

Thanks to the cryptographic hash function, chunks could be evenly distributed across all nodes, even for
severely skewed workloads.  However, all meta chunks generated by a servlet are always stored in its local
chunk storage, as they are not accessed by other servlets.  By keeping the meta chunks locally, it is
efficient to return a primitive object, or to track historical versions.  In addition, servlets may cache the
frequently accessed remote chunks.  When reading \mytree nodes from a \myvalue, request dispatchers
forward \texttt{Get-Chunk} request directly to the chunk storage, without going through the servlet.

\subsubsection{Re-balancing \mytree Construction}
Constructing the \mytree is computation-intensive, which could become a bottleneck for the servlet.  Since servlets
and chunk storages are decoupled when generating and persisting \mytree, an overloaded servlet can
redistribute the tree construction to other servlets.  First, the servlet locks the branch table of the
target key, and forwards the request to another servlet.  
Upon receiving the \myid of the constructed \mytree, the
first servlet embeds it into the \myvalue, updates and finally unlocks the branch table. In summary, unlike
updating the branch table and \myvalue, \mytree construction can be distributed.

\vspace{-2ex}
\section{Fast Development of Efficient Applications}
\label{sec:app}

In this section, we show how \mystore can be exploited for three applications: a blockchain platform, a wiki engine and
a collaborative analytics application. 
We describe how the storage system meets the applications' demands, reduces
development efforts and offers them high performance.

\subsection{Blockchain}
\label{sec:blockchain}

A blockchain system consists of a set of mutually distrustful nodes that together maintain a ledger data structure,
which is made consistent via a distributed consensus protocol.
Previous works have mainly focused on improving consensus protocols which are shown to be a major performance bottleneck~\cite{blockbench}.
The data model and storage component of the blockchain are
overlooked, although there is an increasing demand for performing analytics on blockchain
data~\cite{morgan16,chainalysis,blocksci}.
The blockchain data consists of some global states and transactions
that modify the states. They are packed into blocks linked with each other via cryptographic hash pointers,
forming a chain that ends at the genesis block. In systems that support smart contracts (user-defined
codes), each contract is given a key-value storage to manage its own states separately from the global states,
and the contract accepts transactions that invoke computations on the states.
We refer readers to~\cite{dinh:2018} for a more comprehensive treatment of the blockchain design space.

Although any existing blockchain can be ported to \mystore, here we focus on Hyperledger for two reasons.  First, it is
one of the most popular blockchains with support for Turing-complete smart contracts, making it easy to evaluate the
storage component by writing contracts that stress the storage. Second, the platform targets enterprise applications
whose demands for both data processing and analytics are more pronounced than public blockchain applications like
crypto-currency.  

\subsubsection{Data Model in Hyperledger}

\begin{figure}
\subfigure[Hyperledger (v0.6) datastructure]{
\includegraphics[width=0.48\textwidth]{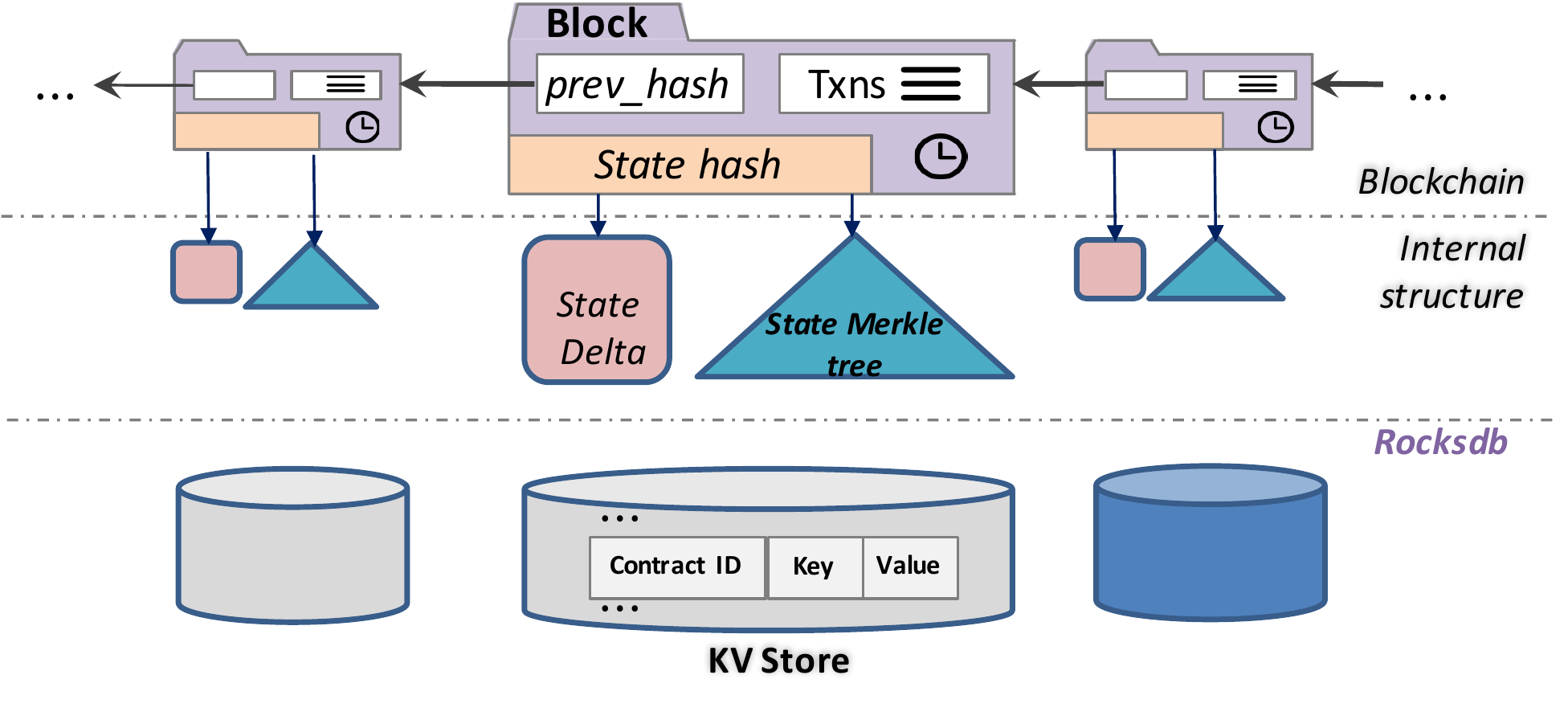}
}
\subfigure[Hyperledger datastructure on \mystore]{
\includegraphics[width=0.48\textwidth]{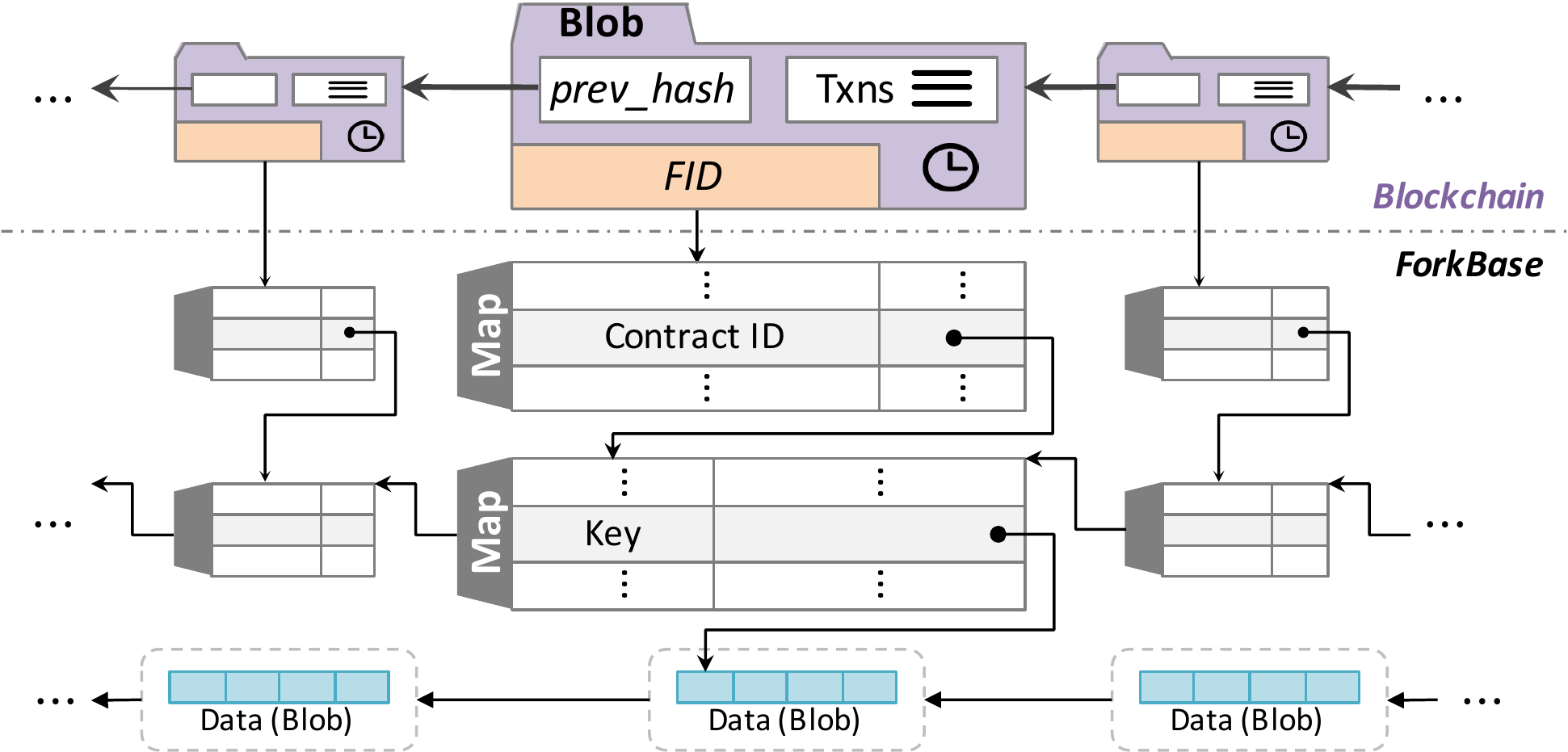}
}
\caption{Blockchain data structures.}
\label{fig:blockchain_design}
\vspace{-1ex}
\end{figure}

Figure~\ref{fig:blockchain_design}(a) illustrates the main data structures in Hyperledger v0.6\footnote{New versions of
Hyperledger, i.e. v1.0 and later, make significant changes to the data model, but they do not fit our definition of a
blockchain system}. The states are protected by a Merkle tree: any modification results in a new Merkle tree; the old
values and old Merkle tree are kept in a separate structure called {\em state delta}.  A blockchain transaction can
issue read or write operations (of key-value tuples) to the states. Only transactions that update the states are stored
in the block. A read operation fetches the value directly from the storage, while write is buffered in a temporary
in-memory data structure. The system batches multiple transactions, then issues a commit  when reaching the desired
number of transactions or when a timer fires. The commit operation first creates a new Merkle tree, then a new state
delta, then a new block, and finally writes all changes to the storage. 

\subsubsection{Blockchain Analytics}
One initial goal of blockchain systems is to securely maintain the states and their histories and thus, the designs are
guided towards achieving tamper evidence and data versioning. 
As blockchain applications gain traction, the massive
volume of data stored on the ledger becomes valuable for analytics. 
Traditional analytical data
management systems, e.g., OLAP databases, achieve high performance by extensive use of indexes and query optimizations.
However, Hyperledger and other blockchains fall short in this respect. 

In this work we consider two representative analytical queries that can be performed on blockchain data. 
The first query is a \emph{state scan}, which returns the history of a given state, i.e.  how the current value comes about. 
The second
query is a \emph{block scan}, which returns the values of the states at a specific block. 
The current Hyperledger data
structures are designed for fast access to the {\em latest} states. 
However, the two above queries require traversing to
the previous states and involve computations with state delta. 
This inefficient query execution is precisely due to the lack
of an index structure. 
We implemented both queries in Hyperledger by adding a pre-processing step that parses all the
internal structures of all the blocks and constructs an in-memory index.  

\subsubsection{Hyperledger on ForkBase }

Figure~\ref{fig:blockchain_design}(b) illustrates how we use \mystore to implement Hyperledger's data structures.  The
key insight here is that an \myvalue fully captures the structure of a block and its corresponding states.
We replace 
Merkle tree and state delta with {\tt Map} objects organized in
two levels. The state hash is now replaced by the version of the first-level {\tt Map} object. This {\tt Map} object contains
key-value tuples where the key is the smart contract ID, and the value is the version of the second-level {\tt Map} object.
This
second-level {\tt Map} contains key-value tuples where the key is the data key, and the value is the version
of a {\tt Blob} object storing the state value. 

One immediate benefit of this implementation is that the code for maintaining data history and integrity becomes
remarkably simple. 
In particular, for only $18$ lines of code that uses \mystore, we eliminate $1918$ lines of code
from the Hyperledger codebase. 
Another benefit is that the data is now readily usable for analytics.  
For state scan query, we simply follow the version number stored in the latest block to get the latest {\tt Blob} object
for the requested key.  From there, we follow $base\_version$ to retrieve the previous values.  For block scan query, we
  follow the version number stored on the requested block to retrieve the second-level {\tt Map} object for this block.
  We then iterate through the key-value tuples and retrieve the corresponding {\tt Blob} objects. 



\subsection{Wiki Engine}

A wiki engine allows users to collaboratively create and edit documents (or wiki entries). Each entry contains
a linear chain of versions.
The wiki engine can be built on top of a multi-versioned key-value storage, in which each entry is mapped to a key and
the entry's content is stored as the associated value. Such a multi-versioned storage can be directly implemented with
Redis~\cite{web:redis}, for instance, using the list data type offered by Redis. More specifically, the wiki entry is of
a list type, and every new version is appended to the list. 

This multi-versioned key-value data model maps naturally into \mystore. Reading and writing an entry are directly
implemented with \texttt{Get} and \texttt{Put} operations on default branches.  Because each version often
changes only small parts of the data, the \texttt{Blob} type is more suitable to represent an entry. Other
meta information, e.g., timestamp, can be stored directly in the \texttt{context} field. 
When
accessing consecutive versions, \mystore can leverage cached data chunks to serve out data more quickly.
Another common operation in a wiki engine, namely \texttt{diff} operation between two versions, is directly
and efficiently supported in \mystore, thanks to the \mytree index.  Finally, \mystore's two-level
partitioning scheme helps alleviate skewed workloads which are common in wiki services due to hot entries. 

\subsection{Collaborative Analytics}

It is becoming increasingly common for a group of scientists (or analysts) to work on a shared dataset, but with
different analysis goals~\cite{bhardwaj:2015,nothaft:2015}.
For example, on a dataset of customer purchasing records,
some analysts may perform customer behavioral analysis, while others use it to improve inventory management. 
At the same
time, the dataset may be continually cleaned and enriched by other analysts.
As the analysts simultaneously work on
different versions or branches of the same dataset, there is a clear need for versioning and fork semantics. 
Decibel~\cite{maddox:2016} and OrpheusDB~\cite{xu:2017} support these features for relational datasets, employing delta-based deduplication techniques.

\mystore has a rich collection of built-in data types, which offer the flexibility to implement many types of
structured datasets.  
Specifically, we implement two layouts for relational datasets: row-oriented and
column-oriented.  
In the former, a record is stored as a \texttt{Tuple}, embedded in a \texttt{Map} keyed
by its primary key.  In the latter, column values are stored as a \texttt{List}, embedded in a
\texttt{Map} keyed by the column name.  Applications can choose the layout that best serves their queries. For instance,
the column-oriented layout is more efficient for applications that perform many analytical queries.

Common operations in collaborative analytics include dataset import and export, dataset transformations,
analytical queries, and version comparisons. In \mystore, accessing large datasets is efficient because only
relevant chunks are fetched to the client. Comparing large datasets via the \texttt{diff} operation is also
efficient, thanks to the \mytree index. 
While other delta-based systems such as Decibel eliminates duplicates
within a single dataset, \mystore deduplication works across multiple datasets, therefore achieving lower storage overhead. 

\begin{table}
  \centering
  \caption{Performance of \mystore Operations.}
  \label{tab:ops_tp}
  \vspace{1ex} 
  \scriptsize
  \begin{tabular}{r|R{1.1cm}|r|R{1.1cm}|r}
    & \multicolumn{2}{c|}{Throughput (ops/sec)} & \multicolumn{2}{c}{Avg. latency (ms)} \\ \cline{2-5}
    & 1KB & 20KB & 1KB & 20KB \\ \hline
    Put-String & 75.0K & 8.3K & 0.24 & 0.9 \\ 
    Put-Blob & 37.5K & 5.7K & 0.28 & 1.0 \\ 
    Put-Map & 35.8K & 4.7K & 0.38 & 1.28 \\ \hline
	Get-String & 78.3K & 56.9K & 0.23 & 0.8 \\ 
    Get-Blob-Meta & 99.7K & 100.4K & 0.16  & 0.17 \\ 
    Get-Blob-Full & 38.4K & 4.9K & 0.62 & 2.9 \\ 
    Get-Map-Full & 38.2K & 5.0K & 0.61 & 3.2 \\ \hline
	Track & 97.8K & 96.0K & 0.16 & 0.17 \\    
    Fork & 113.6K & 109.4K & 0.17 & 0.17 \\ \hline
\end{tabular}
\end{table}

\section{Evaluation}
\label{sec:exp}

We implemented \mystore in about 30k lines of \texttt{C++} code.  
In this section, we
first evaluate the performance of \mystore operations. 
Next, we evaluate the three applications discussed in
Section~\ref{sec:app} in terms of storage consumption and query efficiency. We compare
them against their respective state-of-the-art implementations.

Our experiments were conducted in an in-house cluster with 64 nodes, each of which runs Ubuntu 14.04, and is
equipped with E5-1650 3.5GHz CPU, 32GB RAM, and 2TB hard disk.  
All nodes are connected via 1Gb Ethernet.
For fair comparison against other systems, all servlets are configured with one thread for request execution and
two threads for request parsing.  
Both leaf and index chunk sizes in the \mytree are set to 4KB.

\subsection{Micro-Benchmark}

We benchmark 9 \mystore operations. We deployed one servlet and used multiple clients for sending requests.
Table~\ref{tab:ops_tp} lists the aggregated throughput and average latency measured at 32 clients, with
varying request sizes. 
We observe that large requests achieve higher network throughput --- the product of throughput and request
size --- because of smaller overheads in message parsing. The throughputs of primitive types are higher than those
of chunkable types, due to overhead in chunking and traversing the \mytree. \texttt{Get-X-Meta},
\texttt{Track} and \texttt{Fork} achieve the highest throughputs, regardless of the request sizes. This is
because these operations require no or very small data transfer. The average latencies of different
operations do not vary much, because the latency is measured at the client side, therefore network delays have
major contribution to the final latency.

\begin{table}
  \centering
  \caption{Breakdown of Put Operation ($\mu s$).}
  \label{tab:ops_breakdown}
  \vspace{1ex} 
  \begin{tabular}{r|r|r|r|r}
    & \multicolumn{2}{c|}{String} & \multicolumn{2}{c}{Blob} \\ \cline{2-5}
  	& 1KB & 20KB & 1KB & 20KB \\ \hline
    Serialization & 0.8 & 0.8 & 1.1 & 1.5 \\
    Deserialization & 5.7 & 9.2 & 6.2 & 13.2 \\
    CryptoHash & 8.5 & 56.4 & 9.5 & 80.6 \\
    RollingHash & - & - & 7.5 & 42.2 \\
    Persistence & 10.4 & 60.7 & 10.5 & 93.7 \\
\end{tabular}
\end{table}

Table~\ref{tab:ops_breakdown} details the cost breakdown of the \texttt{Put} operation, excluding the network
cost.  
It can be seen that the main contributor to the latency gap between primitive and chunkable types is 
the rolling hash computations incurred in the \mytree. 

\begin{figure}
  \centering
  \includegraphics[width=0.42\textwidth]{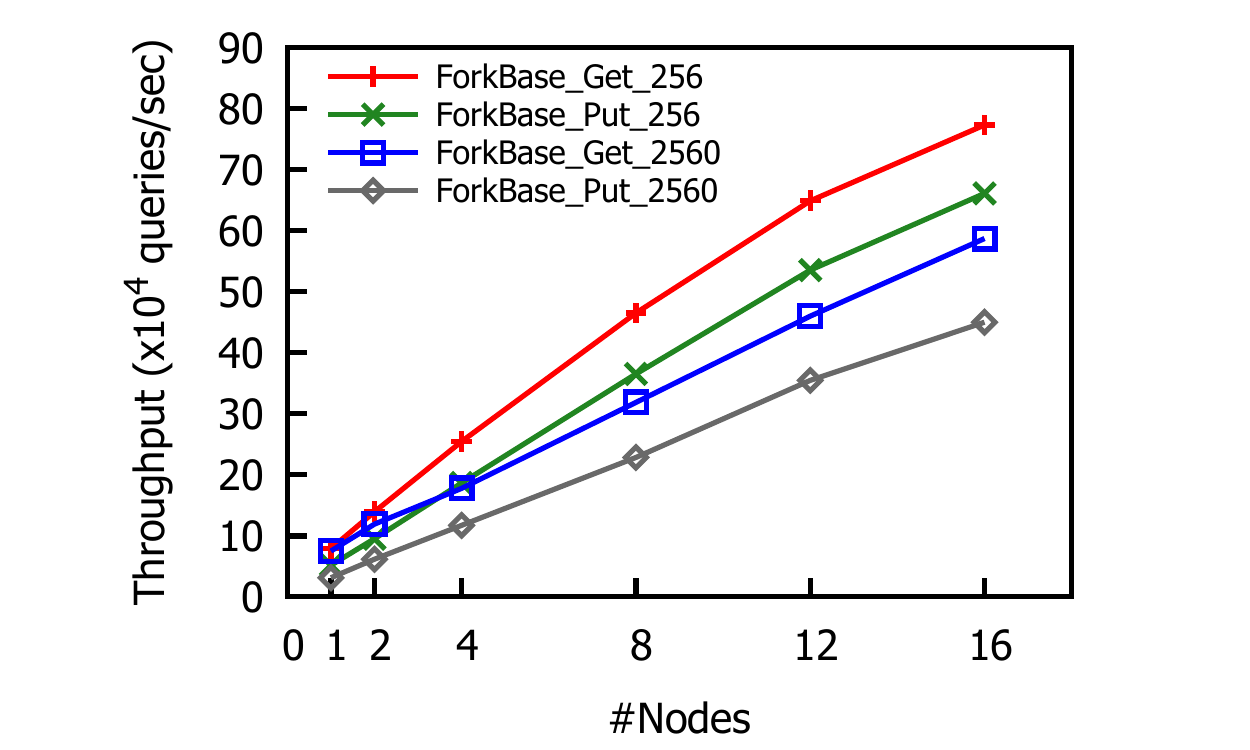}
  \vspace{-1.5ex}
  \caption{Scalability with multiple servlets.}
  \label{fig:exp:scalability}
\vspace{-1ex}
\end{figure}

\begin{figure*}
\centering
\subfigure[Read]{
  \includegraphics[scale=0.55]{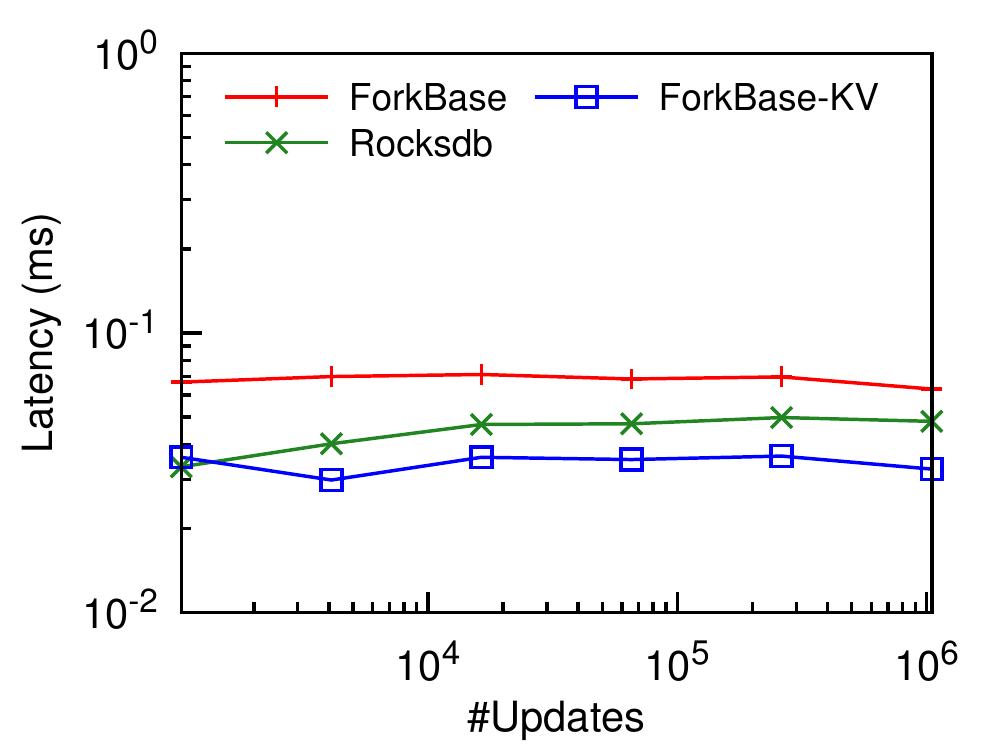}
  \label{fig:blockchain_read}}
\subfigure[Write]{
  \includegraphics[scale=0.55]{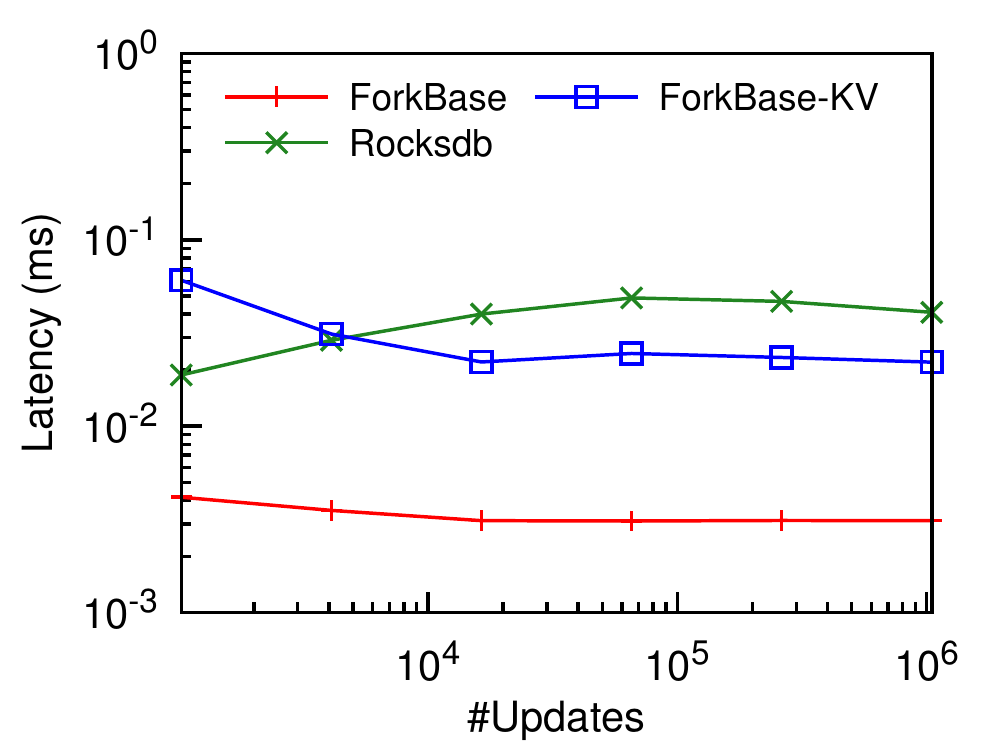}
  \label{fig:blockchain_write}}
\subfigure[Commit]{
  \includegraphics[scale=0.55]{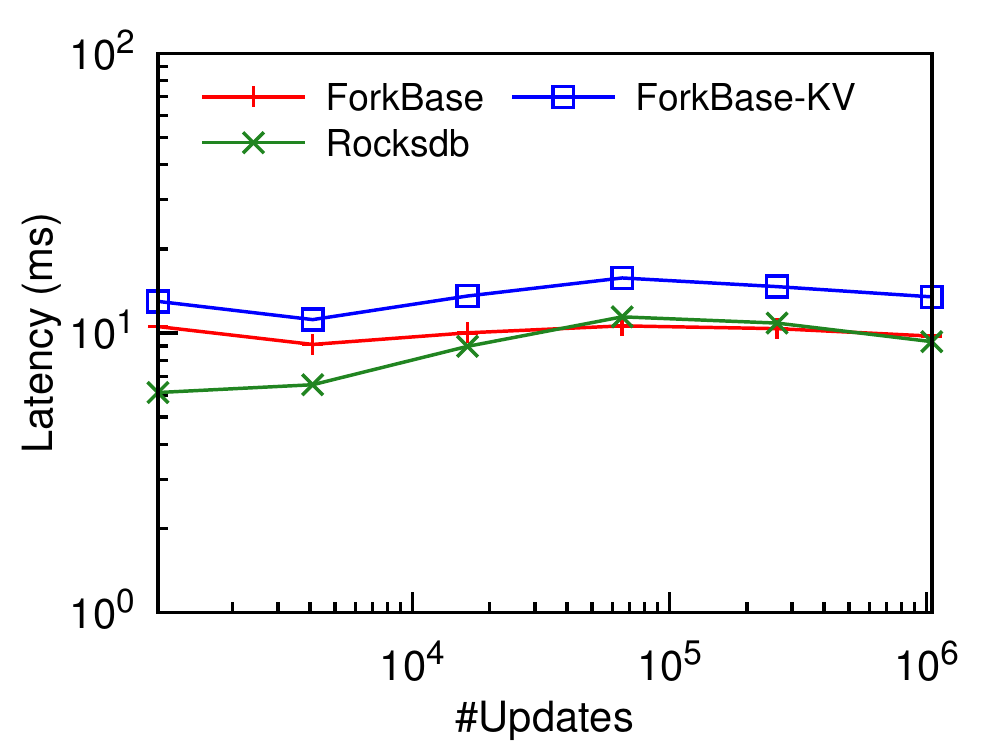}
  \label{fig:blockchain_commit}}
\vspace{-1.5ex}
\caption{Latency of blockchain operations (b=50, r=w=0.5).}
\label{fig:blockchain_operations}
\vspace{-1ex}
\end{figure*}

\begin{figure*}
\centering
\begin{minipage}{.48\textwidth}
  \centering
  \includegraphics[scale=0.6]{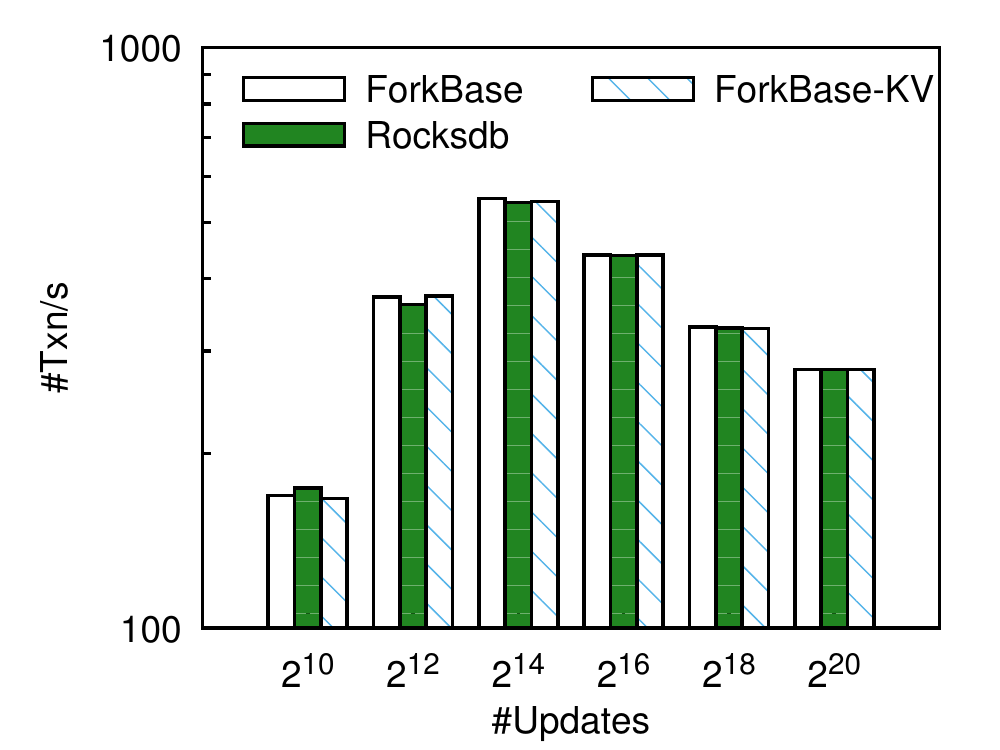}
  \vspace{-1.5ex}
  \caption{Client perceived throughput (b=50, r=w=0.5).}
  \label{fig:blockchain_throughput}
\end{minipage}
\hspace{1em}
\begin{minipage}{.48\textwidth}
  \centering
  \includegraphics[scale=0.70]{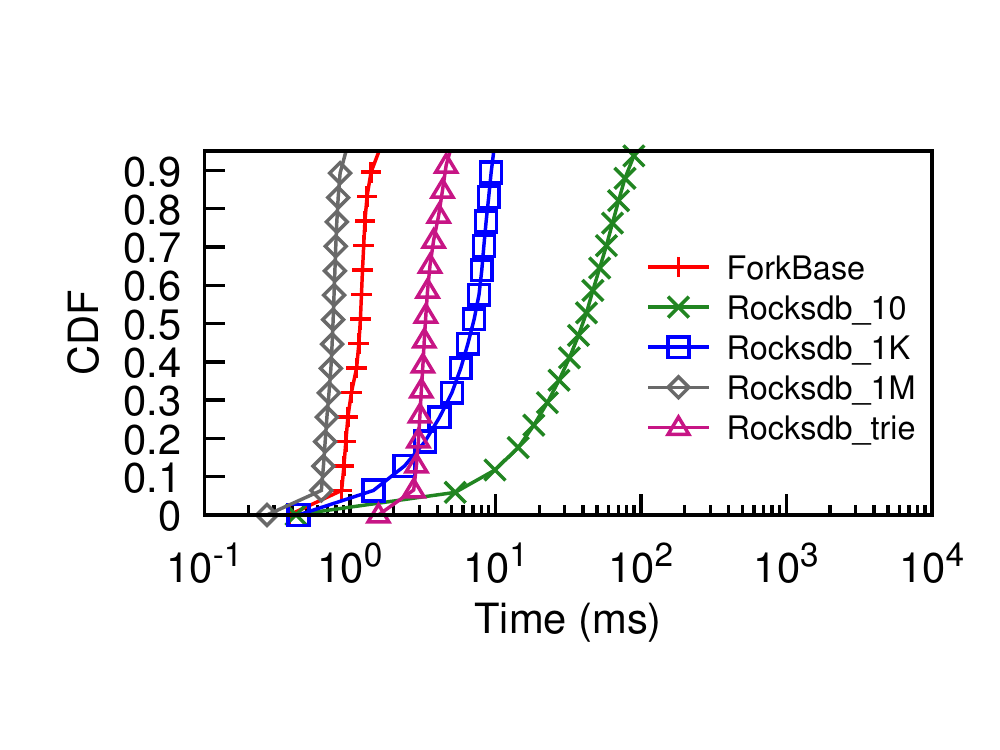}
  \vspace{-3ex}
  \caption{Commit latency distribution with different Merkle trees.}
  \label{fig:commit_latency}
\end{minipage}
\vspace{-1ex}
\end{figure*}

We measured \mystore's scalability by increasing the number of servlets up to 64.
Figure~\ref{fig:exp:scalability} shows almost linear scalability for both \texttt{Put} and \texttt{Get}
operations. 
The fact that \mystore scales almost linearly is expected because there is no communication between the servlets. 

\subsection{Blockchain}
We compare \mystore-backed Hyperledger~\cite{web:hyperledger} with the original implementation using RocksDB, and
also with another implementation that uses \mystore as a pure key-value storage. 
We refer to them as \mystore, Rocksdb
and \mystore-KV respectively. 
We first evaluate how different storage engines affect normal operations of
Hyperledger and the user-perceived performance.  
We then evaluate their efficiency on supporting analytical
queries.  
We used Blockbench~\cite{blockbench}, a benchmarking framework for permissioned blockchains, to
generate and drive workloads.
Specifically, we used the smart contract implementing a
key-value store.  
Transactions for this contract are generated based on YCSB workloads. 
We varied the number
of keys, the number and ratio of read and write operations ($r$ and $w$). Unless stated otherwise, the number of
keys is the same as the number of operations.  
For the blockchain configuration, we deployed one server,
varied the maximum block size $b$ and kept the default values for the other settings. 


\subsubsection{Blockchain Operations}

\hlight{
Figure~\ref{fig:blockchain_operations} shows the $95^{th}$ percentile latency of
blockchain operations, including read, write and commit.
As we can observe, both read and write operations take under $0.1ms$,
two orders of magnitude faster than a commit.
A read only retrieves one key at a time from the storage.
\mystore takes longer than the other two because multiple objects needed to be
retrieved.
\mystore-KV is slightly better than Rocksdb, because the latter stores data in multiple levels (based on Leveldb)
and requires traversing them to retrieve the key.
A write in \mystore simply buffers new value, whereas Rocksdb
and \mystore-KV need to compute temporary updates for the internal structures (Merkle tree and state delta).
This explains why \mystore outperforms the others in write.
}
Though Rocksdb is designed for fast batch commits, \mystore and Rocksdb still have similar latencies, as shown in Figure~\ref{fig:blockchain_commit}.
Both are better than \mystore-KV since using \mystore as a pure key-value store introduces overhead from doing hash computation both inside and outside of the storage layer.

Figure~\ref{fig:blockchain_throughput} shows the overall throughput, measured as the total number of transactions committed to the blockchain per second.
We see no differences in throughput, because the overheads in read, write and commit are relatively small compared to the total time a transaction takes to be included in the blockchain.
In fact, we observe that the cost of executing a batch of transactions is much
higher than that of committing the batch.

\begin{figure*}
\centering
\subfigure[State Scan]{
  \includegraphics[scale=0.55]{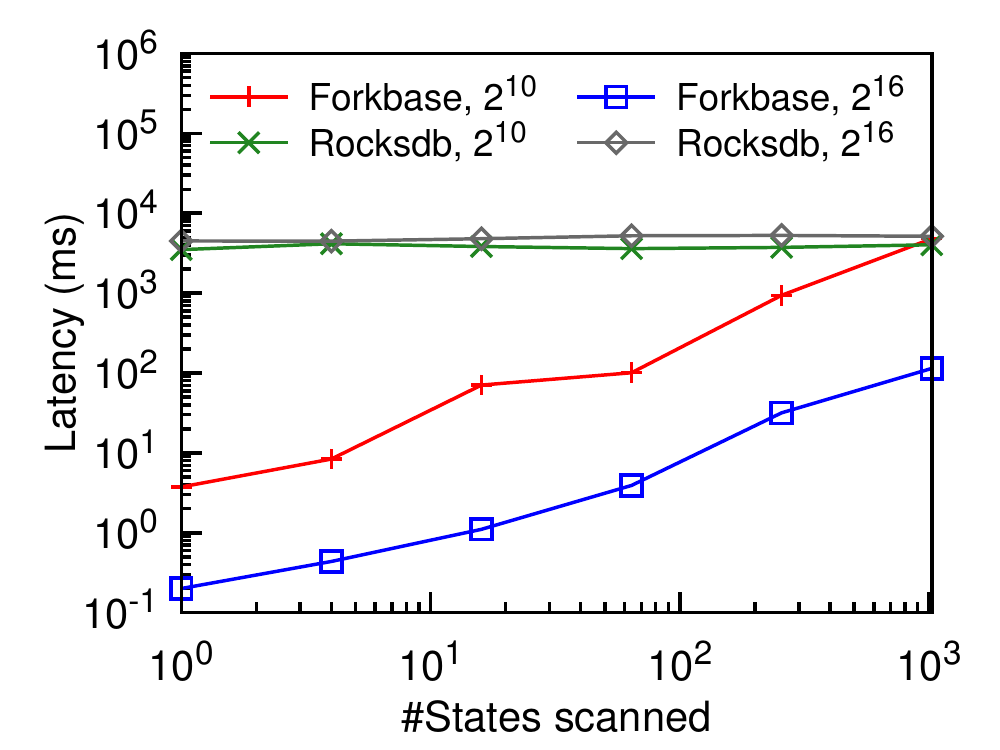}
  \label{fig:blockchain_state_query}}
\hspace{6em}
\subfigure[Block scan]{
  \includegraphics[scale=0.55]{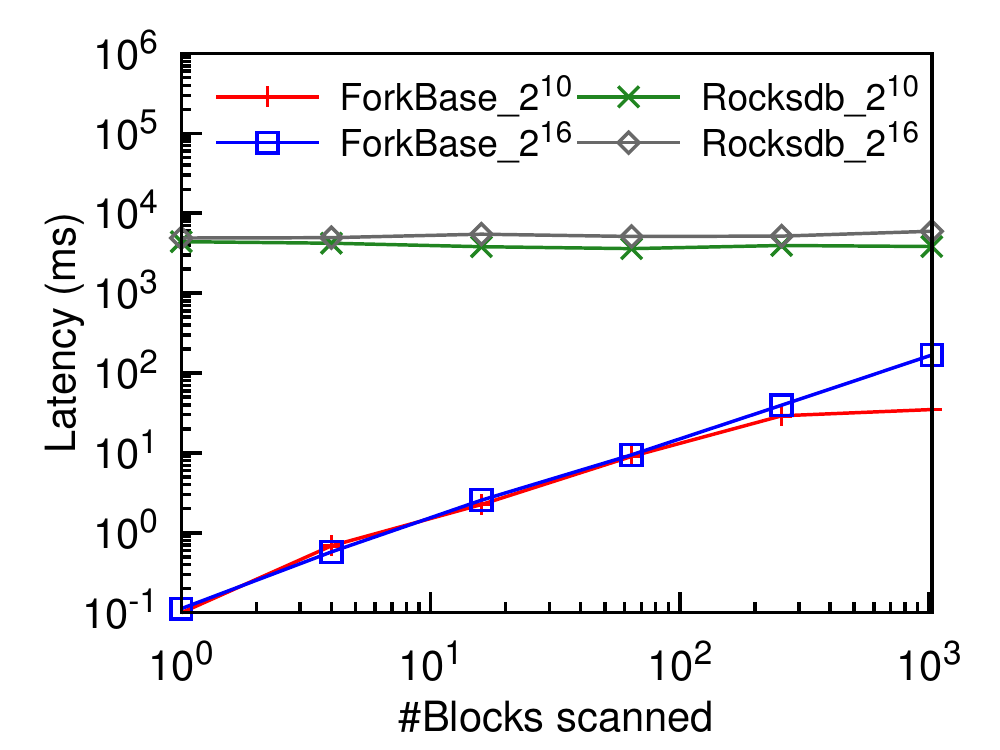}
  \label{fig:blockchain_block_query}}
\vspace{-1.5ex}
\caption{Scan queries. 'X, $2^{y}$' means using storage X, and $2^{y}$ keys.}
\vspace{-1ex}
\end{figure*}

\begin{figure*}
\centering
\subfigure[Throughput]{
  \includegraphics[scale=0.55]{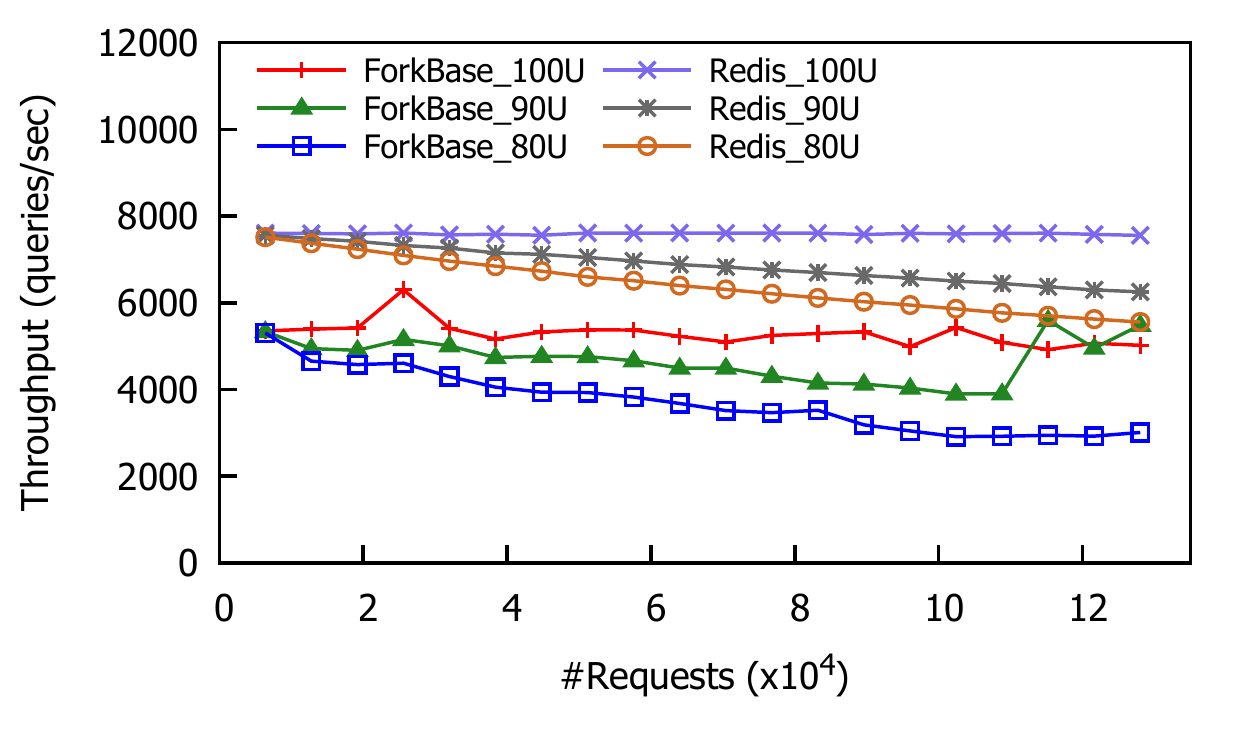}
  \label{fig:exp:wiki_write_tp}}
\hspace{3em}
\subfigure[Storage Consumption]{
  \includegraphics[scale=0.55]{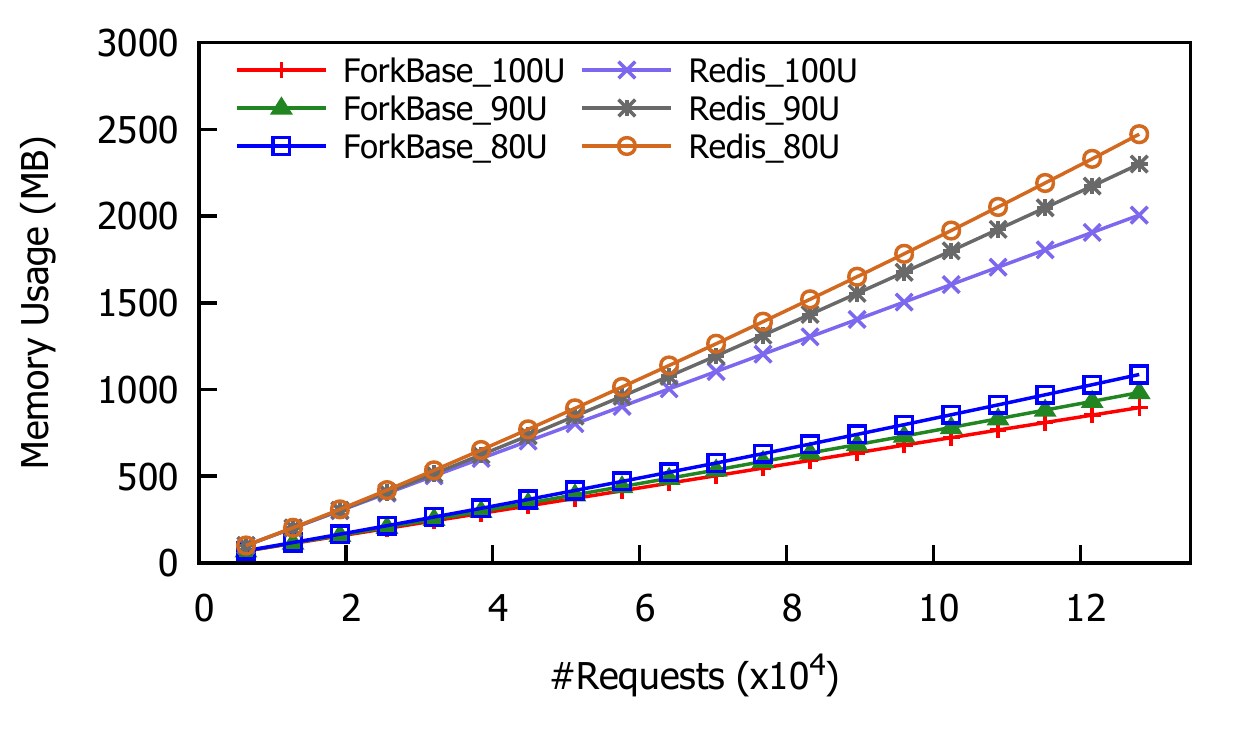}
  \label{fig:exp:wiki_write_sp}}
\vspace{-1.5ex}
\caption{Performance of editing wiki pages.}
\vspace{-1ex}
\end{figure*}

\subsubsection{Merkle Trees}
A commit operation involves updating {\tt Map} objects in \mystore or the Merkle trees in the original
Hyperledger. Hyperledger provides two Merkle tree implementations.  The default option is a bucket tree, in
which the number of leaves is fixed and pre-determined at start-up time, and the data key's hash determines
its bucket number.  The other option is a trie.  Figure~\ref{fig:commit_latency} shows how different
structures affect the commit latency. 
With bucket tree, the number of buckets ($nb = 10$, 1K, 1M) has
significant impact on the commit latency.
With fewer buckets, 
the latency increases and the distribution becomes less uniform.
This is because with more updates, write amplification becomes more severe, which increases the cost of
updating the tree. 
In fact, for any pre-defined number of buckets, the bucket tree is expected to fail to
scale beyond workloads of a certain size. In contrast, \texttt{Map} objects in \mystore scale gracefully by
dynamically adjusting the tree height and bounding node sizes.  The trie structure exhibits low amplification,
but the latency is higher than \mystore because the structure is not balanced, therefore it may require longer
tree traversals during updates.

\subsubsection{Analytical Queries}
We populated the storage with varying numbers of keys and a large number of updates that result in a
medium-size chain of $12000$ blocks. Figure~\ref{fig:blockchain_state_query} compares the performance for
state scan query. 
The x axis represents the number of unique keys scanned per query. For a small number of
keys, the difference between \mystore and Rocksdb is up to 4 orders of magnitudes. 
This is because the cost in
Rocksdb is dominated by the pre-processing phase, which is not required in \mystore. However, this cost
amortizes with more keys, explaining why the performance gap gets smaller. In particular, this gap reduces
to $0$ when the number of unique keys being scanned is the same as the total number of keys in the storage,
since scanning them requires
retrieving all the data from the storage. 


Figure~\ref{fig:blockchain_block_query} shows the performance for block scan query. The x axis represents the block
number being scanned, where x = 0 is the oldest (or first) block. We see a huge difference in performance starting from 4 orders of magnitudes but decreasing with
higher block numbers. The cost in \mystore increases because higher blocks contain more keys to be read. This
also explains why it stops increasing after a certain number of blocks, after which the gap remains at least
two orders of magnitudes regardless of which block being scanned. 

\begin{figure*}
\centering
\begin{minipage}{.48\textwidth}
  \centering
  \includegraphics[scale=0.55]{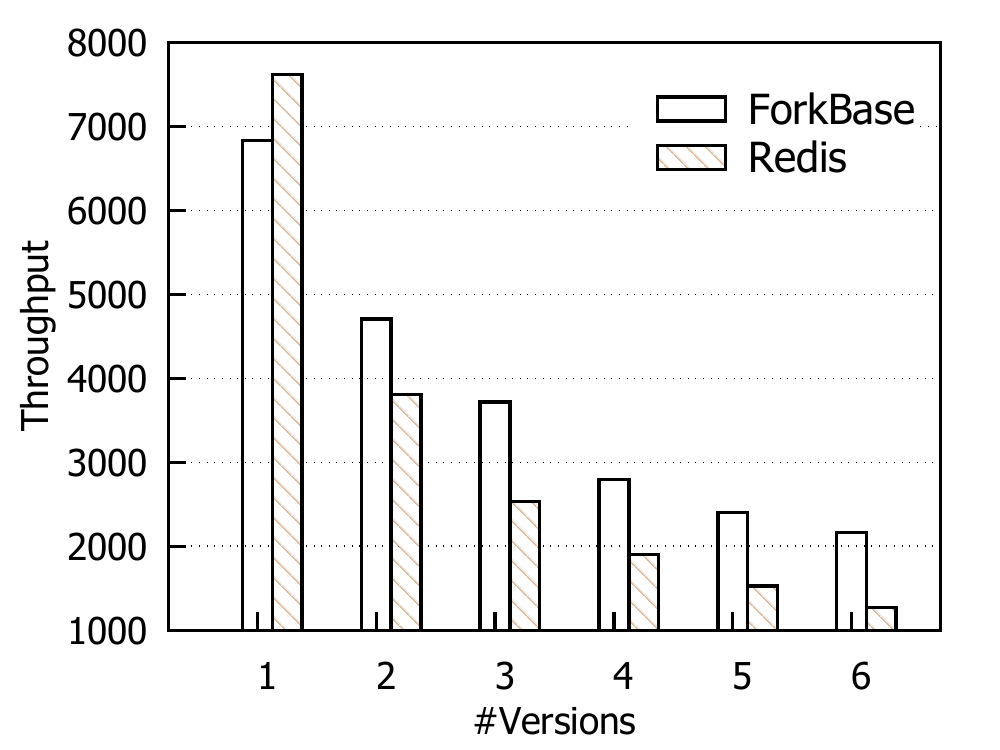}
  \vspace{-1.5ex}
  \caption{Throughput of read consecutive versions of a wiki page.}
  \label{fig:exp:wiki_read}
\end{minipage}
\hspace{1em}
\begin{minipage}{.48\textwidth}
  \centering
  \includegraphics[scale=0.55]{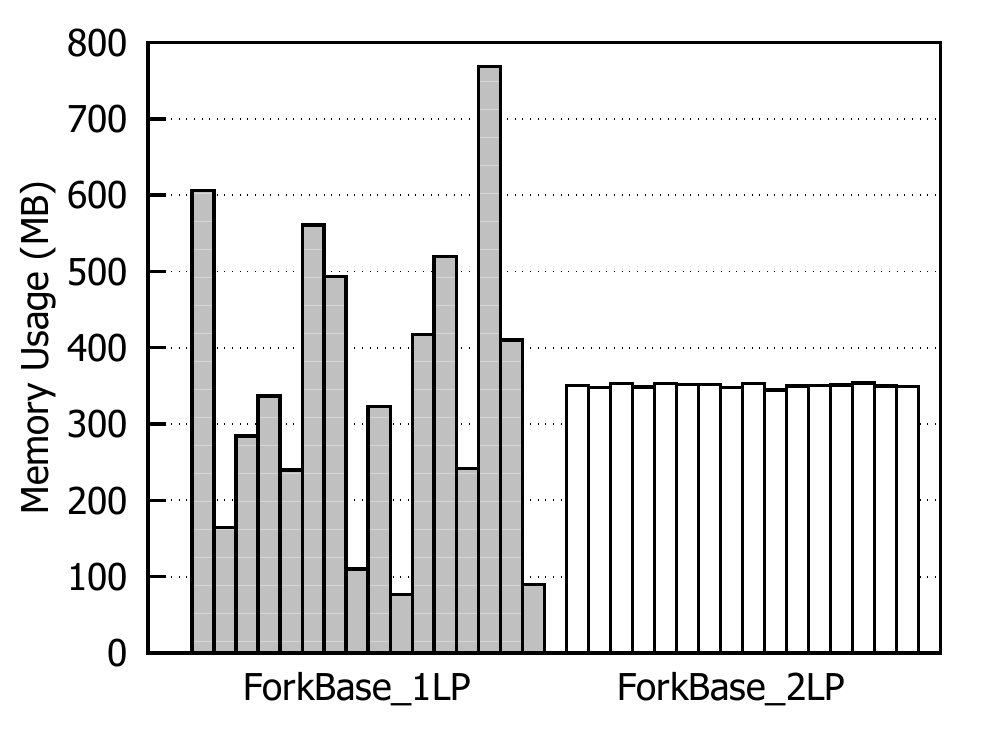}
  \vspace{-1.5ex}
  \caption{Storage size distribution in skewed workloads.}
  \label{fig:exp:wiki_dist}
\end{minipage}
\vspace{-1ex}
\end{figure*}

\begin{figure*}
\centering
\subfigure[Elapsed time]{
  \includegraphics[scale=0.55]{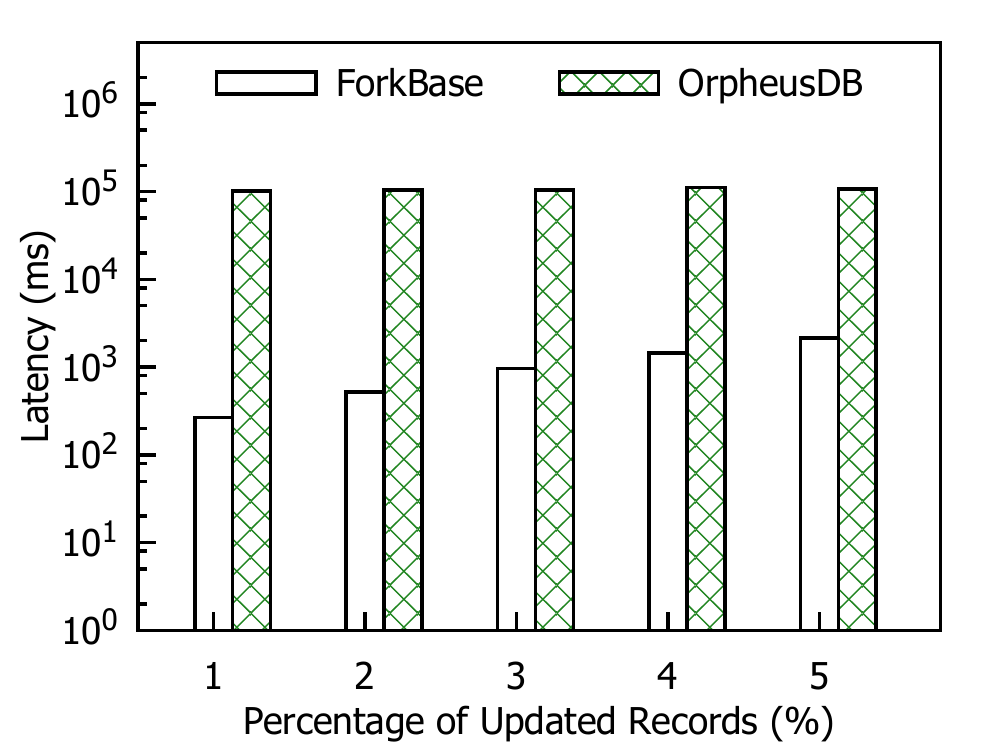}
  \label{fig:exp:ca_mod_time}}
\hspace{7em}
\subfigure[Space Increment]{
  \includegraphics[scale=0.55]{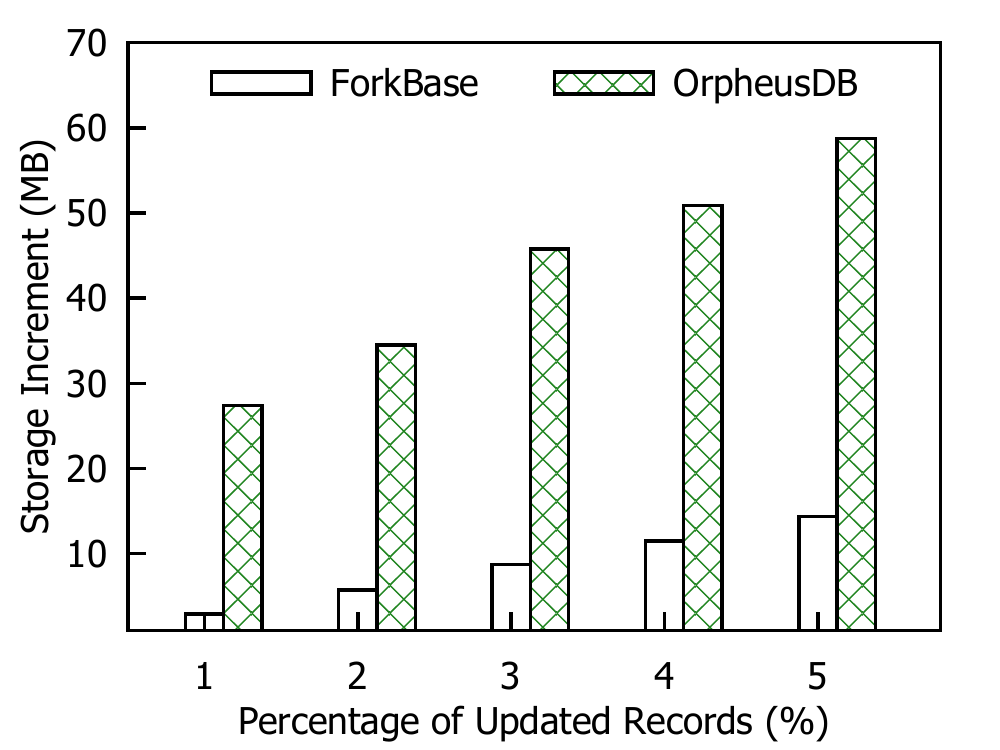}
  \label{fig:exp:ca_mod_space}}
\vspace{-1.5ex}
\caption{Performance of dataset modifications.}
\vspace{-1ex}
\end{figure*}

\subsection{Wiki Engine}

We compare \mystore with Redis, both of which were deployed as multi-versioned wiki engines.  We employed 32
clients on separate nodes to simultaneously edit 3200 pages hosted in the engine.  In each request, a client
loads/creates a random page whose initial size is 15 KB, edits/appends the text, and finally uploads the
revised version.

\subsubsection{Edit and Read Pages}


Figure~\ref{fig:exp:wiki_write_tp} shows the throughput of
editing pages, in which \texttt{xU} indicates the ratio of in-place updates against insertions (\texttt{100U}
means all updates are in-place).  It is expected that Redis outperforms \mystore in terms of write throughput,
since the latter has to chunk the text and build the \mytree.  On the other hand, the chunking overhead is
paid off by the deduplication along the version history.  As shown in
Figure~\ref{fig:exp:wiki_write_sp},
\mystore consumes $50\%$ less storage than Redis,
even though Redis uses compression during data persistence. The performance of reading wiki pages is illustrated in
Figure~\ref{fig:exp:wiki_read}.
It can be seen that
Redis is fast for reading the latest version.
As we track more versions during a single exploration,
\mystore starts to outperform Redis.  The reason is that the data chunks composing a \texttt{Blob} value can
be cached at the clients. When reading an old version, a large number of chunks may have already been cached,
resulting in smaller read latencies.

\subsubsection{Hot Pages}

We deployed a distributed wiki service in a 16-node cluster, and ran a skewed workload (zipf = 0.5).
Figure~\ref{fig:exp:wiki_dist} shows the effect of skewness to storage size distribution. With one layer
partitioning on the page name (1LP), where page content is stored locally, \mystore suffers from imbalance.
The two layer partitioning (2LP) overcomes the problem by distributing chunks evenly among different nodes.

\subsection{Collaborative Analytics}

We compare \mystore with OrpheusDB, a state-of-the-art dataset management system, in terms of their
performance in storing and querying relational datasets. \hlight{For completeness, our comparison contains update
queries, but we note that  OrpheusDB is not designed for efficient checkout and modification operations.}
We used a dataset containing 5 million records loaded from a csv file.
Each record is around 180 bytes in length, consisting of a 12-byte primary key,
two integer fields and other textual fields of variable lengths.
The initial space consumption for this dataset is 927MB in
\mystore and 1167MB in OrpheusDB.

\begin{figure*}
\centering
\subfigure[Diff]{
  \includegraphics[scale=0.55]{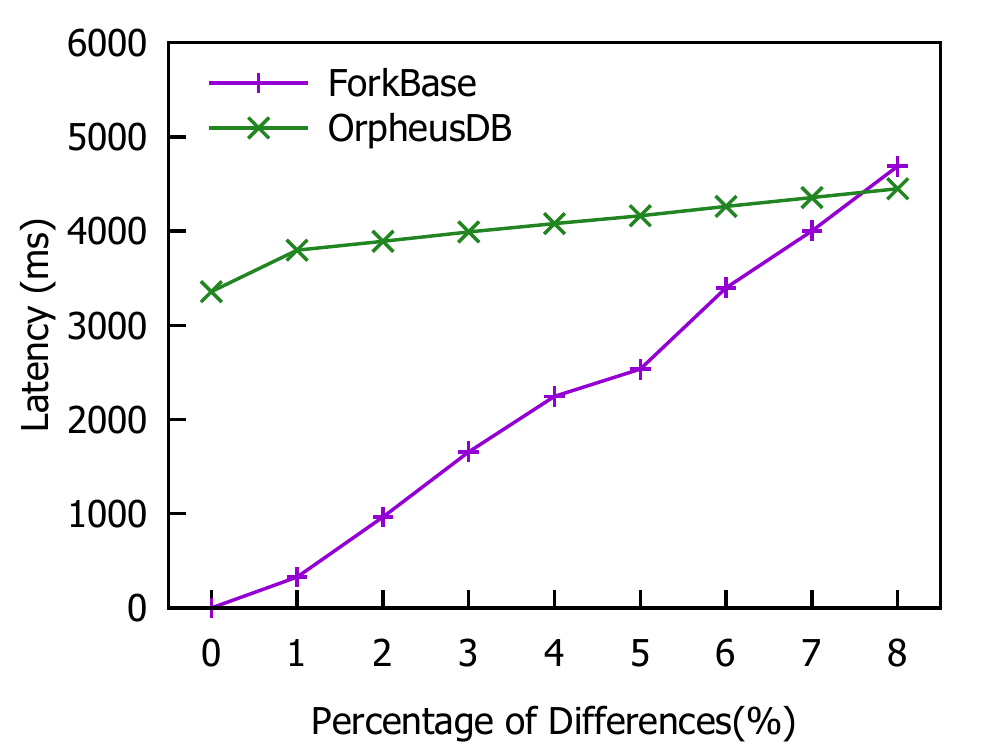}
  \label{fig:exp:ca_query_diff}}
\hspace{6em}
\subfigure[Aggregation]{
  \includegraphics[scale=0.55]{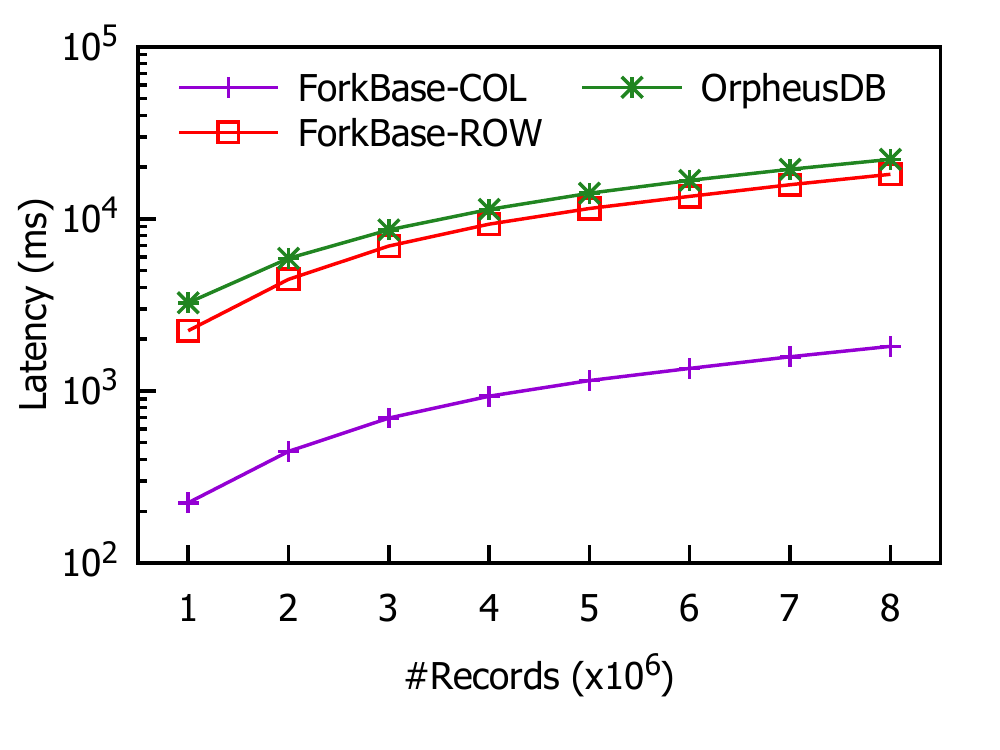}
  \label{fig:exp:ca_query_agg}}
\vspace{-1.5ex}
\caption{Performance of querying datasets.}
\vspace{-1ex}
\end{figure*}

\hlight{
\subsubsection{Dataset Modification}
The dataset maintained by OrpheusDB is materialized as a table and can be easily manipulated through standard
SQL queries.  \mystore uses built-in methods to implement the same table abstraction. 
Figure~\ref{fig:exp:ca_mod_time} shows the latency of
dataset modification for both systems.  As OrpheusDB is not designed for fast data updates, \mystore
outperforms it by two orders of magnitude. The performance gap is due to two factors.  First, during
checkout, OrpheusDB reconstructs a working copy from sub-tables, whereas \mystore only returns a handler and
defer fetching relevant chunks. Second, during commit, there is less data to be stored,
as can be seen from Figure~\ref{fig:exp:ca_mod_space}.  OrpheusDB consumes $3 \times$ more space than
\mystore from newly created sub-tables.  Thanks to fine-grained deduplication, \mystore only needs to
commit a small number of chunks.
}

\subsubsection{Version Comparison}

Figure~\ref{fig:exp:ca_query_diff} shows the cost in comparing two dataset versions with a varying degree of
differences. OrpheusDB's cost is roughly consistent, because the storage maintains a vector of record-version mapping
for each dataset version, and it relies on full vector comparison to find the differences.  On the contrary,
  \mystore's cost is low for small differences, because \mystore can quickly locate them using
  the \mytree.  However, the cost increases when the differences are large, because \mystore has to traverse
  more tree nodes.

\subsubsection{Analytical Queries}

Figure~\ref{fig:exp:ca_query_agg} compares the performance of aggregation queries on the numerical fields.
For \mystore, both row and column layouts were used. 
It can be seen that row-oriented \mystore and OrpheusDB have similar performance, whereas column-oriented
\mystore has $10 \times$ better performance.  The gap is due to the physical layouts. More specifically,
extracting fields is more expensive in row-oriented than in column-oriented layouts. This shows that
\mystore's flexible data model offers opportunities to fine-tune application performance.
Finally, we note that OrpheusDB supports other advanced analytics, e.g., join queries, thanks to its underlying RDBMS. Nevertheless, with additional engineering effort, it is possible to extend \mystore with richer query
functionalities by adding them to the view layer.

\section{Conclusions}
\label{sec:conclu}

There are requirements from modern applications that have not been well addressed in existing storage systems.  We
identified three common properties in blockchain and forkable applications: data versioning, fork semantics
and tamper evidence. 
We discussed the values of a unified storage engine that offers these properties off the
shelf. 
We designed and implemented \mystore that embeds the above properties and is able to deliver better
performance than ad-hoc, application-layer solutions.  By implementing three applications on top of \mystore,
we demonstrate that our storage makes it easy to express application requirements, thereby reducing
development efforts.  
We showed via experimental evaluation that \mystore is able to deliver better
performance than state-of-the-art in terms of storage consumption, query efficiency and coding complexity.  We
believe that \mystore's unique properties are key enablers for building emerging
applications such as blockchain and collaborative analytics.


\bibliographystyle{abbrv}
\bibliography{main-bibliography}

\end{document}